\documentclass[aps,pra,floatfix,superscriptaddress,notitlepage,twocolumn]{revtex4-2}

\usepackage[T2A]{fontenc}
\usepackage[utf8]{inputenc}
\usepackage[english]{babel}
\usepackage[colorlinks=true, allcolors=blue]{hyperref}
\usepackage{bm}
\usepackage{graphicx}
\usepackage{amsmath}
\usepackage{amssymb}
\usepackage{array}
\usepackage{multirow}
\usepackage{float}
\usepackage{braket}
\usepackage{feynmp-auto}
\usepackage{tikz-feynman}
\usetikzlibrary{decorations.pathmorphing}
\tikzfeynmanset{compat=1.1.0}

\newcommand{\ii}{\mathrm{i}} 
\usepackage{array}
\usepackage{diagbox}

\begin{document}

\title{Fully relativistic calculation of thermal shifts in spectra of hydrogen molecular ion}
\author{M.~A.~Reiter} 
\email[E-mail:]{ mikh.reiter@gmail.com}
\affiliation{Department of Physics, St. Petersburg State University, Universitetskaya nab. 7/9, 199034 St. Petersburg, Russia}

\author{A.~A.~Anikin}
\affiliation{Department of Physics, St. Petersburg State University, Universitetskaya nab. 7/9, 199034 St. Petersburg, Russia}
\affiliation{D. I. Mendeleev Institute for Metrology, St. Petersburg, 190005, Russia}

\author{A.~A.~Kotov}
\affiliation{Department of Physics, St. Petersburg State University, Universitetskaya nab. 7/9, 199034 St. Petersburg, Russia}

\author{D.~A.~Glazov}
\affiliation{School of Physics and Engineering, ITMO University, Kronverkskiy pr. 49, 197101 St. Petersburg, Russia}
\affiliation{Petersburg Nuclear Physics Institute named by B.P. Konstantinov of NRC ``Kurchatov Institute'', Orlova roscha 1, 188300 Gatchina, Leningrad region, Russia}

\author{T.~A.~Zalialiutdinov}
\affiliation{Department of Physics, St. Petersburg State University, Universitetskaya nab. 7/9, 199034 St. Petersburg, Russia}
\affiliation{Petersburg Nuclear Physics Institute named by B.P. Konstantinov of NRC ``Kurchatov Institute'', Orlova roscha 1, 188300 Gatchina, Leningrad region, Russia}

\author{D.~A.~Solovyev}
\affiliation{Department of Physics, St. Petersburg State University, Universitetskaya nab. 7/9, 199034 St. Petersburg, Russia}
\affiliation{Petersburg Nuclear Physics Institute named by B.P. Konstantinov of NRC ``Kurchatov Institute'', Orlova roscha 1, 188300 Gatchina, Leningrad region, Russia}

\begin{abstract}

Within a fully relativistic framework, we discuss the one-loop self-energy correction, generalized to the presence of external blackbody radiation, for a bound electron in the hydrogen molecular ion. The calculations rely on the thermal part of the photon propagator from finite-temperature quantum electrodynamics. The ADKB method, which extends the dual kinetic balance condition to axially symmetric systems, is adopted together with a B-spline-based trial function. The results are obtained within the Born-Oppenheimer adiabatic approximation. Thermal shift for the ground term of the hydrogen molecular ion is computed as a function of the internuclear distance. Including this correction, we present the resulting shifts of the vibrational levels. The analysis focuses on the dynamical effect, a comparison is made with data available in the current literature.
\end{abstract}

\maketitle

\section{Introduction}

In recent years, investigations of thermal environment effects on atomic systems have become an essential component of precision spectroscopic experiments. First detected in the late 1970s \cite{GC}, this effect was rapidly interpreted within the framework of quantum mechanics \cite{farley}. Studying physical phenomena under laboratory conditions primarily requires accounting for an external photon field, which is determined by the temperature and described by the Planck frequency distribution \cite{FW,Abr,Das,Kleinert}. Because such thermal radiation is isotropic, its effects appear only in second‑order perturbation theory. When considering the dipole interaction with an external electric field, a quadratic Stark shift can also arise in the thermal case. Here, the squared field strength can be derived from the intensity of the surrounding thermal photon environment \cite{farley}.

However, in the case of a photon flux, in addition to the electric component of intensity, there is also a magnetic component, which can likewise be obtained from the Planck distribution. Inclusion of the magnetic component has an effect already on the fine‑ and hyperfine‑structure splitting \cite{Itano,PhysRevA.110.043108}. The magnetic thermal effect is commonly associated with the Zeeman shift. It is worth noting that the diamagnetic interaction also falls into this category of phenomena \cite{Beloy_2025}.

Due to the complexity of performing an exact analytical calculation, estimates of the thermal Stark effect have been carried out within various approximations. One of these is the static limit: under the assumption that the frequency of the external thermal photon field is sufficiently small, this approximation is quite relevant at room temperatures (i.e., under laboratory conditions). Another approximation is the Bates–Damgaard method \cite{farley}, employed for the approximate description of the dynamic atomic polarizability. The latter, however, does not always provide satisfactory accuracy. In particular, in the works \cite{Reiter_2026,Bobylev_2026} significant discrepancies in the results for fine-structure states were found. It is noteworthy that more recent studies have performed accurate numerical calculations of the dynamic quantities directly, without employing the aforementioned approximations.

Quantum-mechanical approaches to evaluating thermal shifts, including Stark, Zeeman, and diamagnetic contributions, remain the primary means of accounting for blackbody radiation in the spectral response of atomic systems. Of particular interest are atoms, molecules, and their ions that are either in use or under consideration as candidates for atomic clock applications, see, e.g., \cite{AtCl-Sr,Saf-Nat,Kozlov-RevModPhys}. The inherent complexity of many‑electron systems in quantum mechanics is the principal motivation for adopting second‑order perturbation theory~\cite{Sahoo-AtCl,chakraborty2023high}. The perturbation operator, which accounts for the interaction with an external thermal photon field, can nevertheless be systematically extended to many‑body systems using established computational techniques. The quadratic thermal Stark effect is usually treated by expanding the dynamic polarizability as a series in a small frequency parameter (at room temperature). As a result, the dominant contribution comes from the static polarizability of the atomic system, while the higher-order terms account for dynamic corrections \cite{PhysRevA.51.3883,Walls2001,Beloy,Bai_2020,PhysRevLett.132.233201}. Presently, these thermal dynamical corrections are considered among the primary sources of experimental uncertainty in atomic and nuclear clocks~\cite{PhysRevLett.134.113801,Yu2026}.

Recently, the finite-temperature quantum electrodynamics (TQED) formalism has been proposed for calculating thermal shifts in atoms and atomic systems \cite{PRA_2015,S_2020}. In particular, it has been shown that within this framework, the well-known Stark shift emerges as the leading-order contribution from the one-loop self-energy correction of a bound electron in the nonrelativistic approximation. The extension beyond the nonrelativistic limit was carried out very recently in \cite{Lopez_2025}, which demonstrated how the thermal Zeeman shift, shifts arising from higher multipoles, and those due to diamagnetic interactions appear within the one-loop self-energy correction. Demonstrating a clear advantage of the TQED approach, relativistic corrections appear at the same order in temperature as the Stark and Zeeman effects. Moreover, the finite-temperature QED formalism enables a rigorous derivation of thermal-radiation-induced corrections to physical observables such as the g-factor and hyperfine structure \cite{ZGSg_2022,ZGS_2022,ZKS_2023}. Given the rigor of this theory, which reduces to the Feynman–Dyson diagrammatic technique, it becomes possible to identify explicit deficiencies of the phenomenological approach in describing the influence of thermal radiation on the spectra of atomic systems \cite{ZS-LR}.

A compelling alternative to atomic clocks is the proposal to build molecular clocks \cite{PhysRevLett.113.023004}. In this context, the hydrogen molecular ion and its deuterated counterpart hold a distinct advantage, as the three-body problem can now be solved with unprecedented accuracy, see, for example, \cite{Drakebook,Korobov_2000,Pachucki:pra:2017,Alighanbari2025} and many others. Unlike conventional atomic clocks, which rely on electronic transitions, molecular clocks have been proposed primarily as ultra-sensitive probes for determining fundamental physical constants, in particular the proton-to-electron mass ratio \cite{Karr_2025}. Compared to atoms, molecules offer additional degrees of freedom - vibration and rotation. Precision measurements based on vibrational transitions are expected to achieve a relative accuracy of $10^{-18}$ \cite{Schiller2022,Schenkel_2024}.

In view of the planned accuracy of molecular clocks based on the H$_2^+$ ion and its heavier analogues, it is of interest to determine the thermal shifts for vibrational levels. Significant suppression of these shifts is nevertheless expected through measurements employing a specially chosen combination of frequencies \cite{Karr_2014}. In light of the proposed experimental technique, the corresponding theoretical calculations become especially important. As described above, calculating the thermal Stark effect amounts to determining the static (leading-order) polarizability of the system in question. Such calculations have been reported in a fairly broad body of literature \cite{Tsogbayar_2009,Schiller_P,Bakalov2015,Aznabayev_2017}. Typically, such calculations are carried out using the well-established variational approach within the frameworks of nonrelativistic quantum electrodynamics. In the present paper, we address the same problem but within a fully relativistic approach \cite{Reiter_2026}. Specifically, within the Born-Oppenheimer approximation, we solve the Dirac equation with a two-center potential that depends parametrically on the internuclear distance. The resulting energies and wave functions, obtained as functions of the internuclear distance, are then employed to compute the thermal one-loop self-energy correction of a bound electron \cite{PRA_2015}. This circumstance allows us to avoid explicitly isolating the dipole polarizability and corresponding corrections, and instead to perform a careful integration over the frequency of the thermal photon background \cite{Lopez_2025}.

The paper is organized as follows. In the next section, we briefly present the basis of the applied theory and the calculation method. Then, we present the numerical results for the adiabatic potential curves (for the ground state and the first three excited terms) and for the ground state thermal one-loop self-energy (SE) correction to the calculated molecular terms, see section~\ref{NR}. From the adiabatic potentials obtained, the vibrational energy levels are constructed by solving the Schr\"{o}dinger equation with potential energy curves (PECs) that include the internuclear interaction $Z_1Z_2/R$. The thermal shift, obtained in section~\ref{NR} as a function of $R$, is then averaged over the resulting vibrational wave functions, giving the thermal shift of the vibrational spectrum, see section~\ref{VTS}. Discussion and conclusions are given in the last section of the paper. The relativistic units are used throughout the paper: $m_e=\hbar=c=1$.

\section{Theory and methods}
\label{TM}
\subsection{Thermal one-loop self-energy correction}
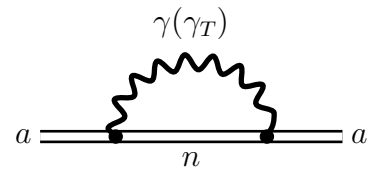
\begin{figure}[h]
\centering
\begin{tikzpicture}
  \coordinate (i) at (0,0);
  \coordinate (f) at (4.0,0);
  \coordinate (a) at (1.0,0);
  \coordinate (b) at (3.0,0);
  
  \draw [double, double distance=3pt, line width=1pt, draw=black] 
        (i) -- (a) -- (b) -- (f);
  
  \filldraw [black] (a) circle (2.5pt);
  \filldraw [black] (b) circle (2.5pt);
  
  \draw [decorate, decoration={snake, amplitude=1mm, segment length=3.5mm}, 
         line width=2pt, draw=black] 
        (a) .. controls (1.0, 1.2) and (3.1, 1.2) .. (b);
  
  \node at (2.0, 1.5) {\large \(\gamma(\gamma_T)\)};
  \node at (i) [left] {\large\(a\)};
  \node at (2.0, -0.3) {\large\(n\)};
  \node at (f) [right] {\large\(a\)};
\end{tikzpicture}
\caption{The one-loop self-energy diagram of a bound electron. Here, the double line represents a bound electron in the nuclear field for an arbitrary atomic state $a$, while the wavy line denotes a photon. The photon is labeled $\gamma$ in the ordinary (zero-temperature) case and $\gamma_T$ in the thermal case.}
\label{fig:1}
\end{figure}

In the framework of finite-temperature quantum electrodynamics, the thermal shift of an atomic level is described by the one-loop self-energy correction of a bound electron \cite{PRA_2015}. In this approach, the effect is determined by the Feynman–Dyson diagram shown in Fig.~\ref{fig:1}, where the photon propagator consists of two contributions - the zero-temperature and the finite-temperature parts.

In the context of the QED $S$-matrix theory \cite{AB,A-B-QED,LabKlim}, the contribution of the diagram depicted in Fig.~\ref{fig:1} reduces to
\begin{eqnarray}
\label{1}
\Delta \varepsilon_a = \langle a | \gamma^0  \Sigma(\varepsilon_a) | a \rangle = \frac{i}{2 \pi} \int\limits_{-\infty}^{\infty} d \omega \sum_n \frac{ \langle a n | I(\omega) | n a \rangle}{\varepsilon_a - \omega - \varepsilon_n^-},
\end{eqnarray}
where
\begin{eqnarray}
\label{2}
I(\omega, \mathbf{r}_1, \mathbf{r}_2) = e^2 \alpha^\nu \alpha^\mu D_{\mu \nu}(\omega, \mathbf{r}_{12}) 
\end{eqnarray}
$\hat{\omega} = \sqrt{\omega^2+i 0} \sim |\omega|, \varepsilon_n^- = \varepsilon_n -\ii 0$, \textcolor{black}{$\boldsymbol{r}_{12}$}$\equiv \mathbf{r}_1 - \mathbf{r}_2$ and $\pmb{\alpha}$ denotes the Dirac $\alpha$-matrix. \textcolor{black}{Here and throughout roman style, $\rm{r}_i$, is used for four-vectors, and italic style, $r_{12} = |\mathbf{r}_{12}|$, for scalars.} Recently, in \cite{PRA_2015}, the photon propagator \textcolor{black}{in the coordinate space} $D_{\mu\nu}(\textcolor{black}{\rm{r}_1, \rm{r}_2})$ was found to take the form:
\begin{eqnarray}
\label{3}
D_{\mu\nu}(\rm{r}_1, \rm{r}_2) = \frac{g_{\mu\nu}}{2\pi\ii r_{12}}\int\limits_{-\infty}^\infty d\omega\, e^{\ii|\omega| r_{12}-\ii \omega(t_1-t_2)}
\\
\nonumber
- \frac{g_{\mu\nu}}{\pi r_{12}}\int\limits_{-\infty}^\infty d\omega\, n_\beta(|\omega|)\sin{|\omega|r_{12}} \textcolor{black}{e^{-\ii \omega(t_1-t_2)}},
\end{eqnarray}
The calculations are performed in the Feynman gauge~\cite{S_2020}. The metric tensor is taken with the Euclidean signature $(+---)$, while the thermal photon distribution is given by the Planck function $n_\beta = (\exp(\omega/k_{\rm B}T)-1)^{-1}$, where $k_{\rm B}$ is the Boltzmann constant and $T$ is the temperature (in kelvin).

In Eq.~(\ref{3}), the first term corresponds to the standard zero‑temperature contribution, whereas the second term accounts for thermally induced effects. Owing to the additivity of the quantity, finite‑temperature effects can be treated separately. The Planck distribution ensures that no ultraviolet divergences appear in the thermal case, so the result is fully regular \cite{ZS_2023,Lopez_2025,Reiter_2026,Bobylev_2026} and gauge invariant \cite{S_2020}. Hence, the energy shift for an arbitrary state $a$ (the dynamical Stark effect in leading order) studied here can be written as
\begin{eqnarray}
\label{4}
\Delta \varepsilon_a^\beta = \frac{\alpha}{\pi} \int\limits_{0}^{\infty} d \omega \; n_\beta ({\omega}) \sum_{n} \left[\frac{ \langle a n | \frac{1 - \pmb{\alpha}_1\pmb{\alpha}_2}{r_{12}}\sin {\omega} r_{12 } | n a \rangle}{\varepsilon_a + \omega - \varepsilon_n^-}
\right.
\nonumber
\\
\left.
+\frac{ \langle a n | \frac{1 - \pmb{\alpha}_1\pmb{\alpha}_2}{r_{12}}\sin {\omega} r_{12 } | n a \rangle}{\varepsilon_a - \omega - \varepsilon_n^-}\right]
\\
\nonumber
\equiv \frac{\alpha}{\pi} \int\limits_{0}^{\infty} d \omega \; n_\beta ({\omega}) \sum_{n,\pm} \frac{ \langle a n | \frac{1 - \pmb{\alpha}_1\pmb{\alpha}_2}{r_{12}}\sin {\omega} r_{12 } | n a \rangle}{\varepsilon_a \pm \omega - \varepsilon_n^-}.
\label{eq:se_t}
\end{eqnarray}
In the above expression, we eliminated the modulus sign present in the photon propagator, employed $e^2 = \alpha$ in relativistic units, and recast the expression into a frequently used equivalent form (last line). The standard notation for the bound state $a$ and the intermediate states $n$ is also adopted, employing bra and ket vectors. The summation over intermediate states runs over the complete Dirac spectrum, including the continuum as well as the negative-energy states.

For further calculations, we employ a partial-wave decomposition
\begin{eqnarray}
    \label{5}
\frac{\sin \omega r_{12}}{4 \pi \omega r_{12}} =  \sum_{l=0}^{\infty} \sum_{m=-l}^{l} j_l(\omega r_<) j_l(\omega r_>) Y_{lm}^* (\hat{r}_1) Y_{lm} (\hat{r}_2),
\end{eqnarray}
where $r_>$ and $r_<$ denote the larger and the smaller of the vector lengths $r_1$ and $r_2$, respectively. For the spherical functions, the standard notation $Y_{lm}$ with orbital angular momentum $l$ and its projection $m$ is used \cite{Varsh}. The dependence on the solid angle is indicated in the argument of the functions as $\hat{r}_1=\bm{r}_1/r_1, \hat{r}_2=\bm{r}_2/r_2$, and $j_l$ is the spherical Bessel function.

\subsection{ADKB method}
\label{ADKB}
Focusing here on systems of two nuclei (heavy particles) and one bound electron, we can numerically solve the Dirac equation with the corresponding two-center potential. Note that such systems are not necessarily equilibrium coumpounds (having an equilibrium internuclear distance). Hence, we shall refer to them as one-electron quasi-molecules. 

\begin{figure}[hbtp]
\centering
\includegraphics[width=0.8\columnwidth]{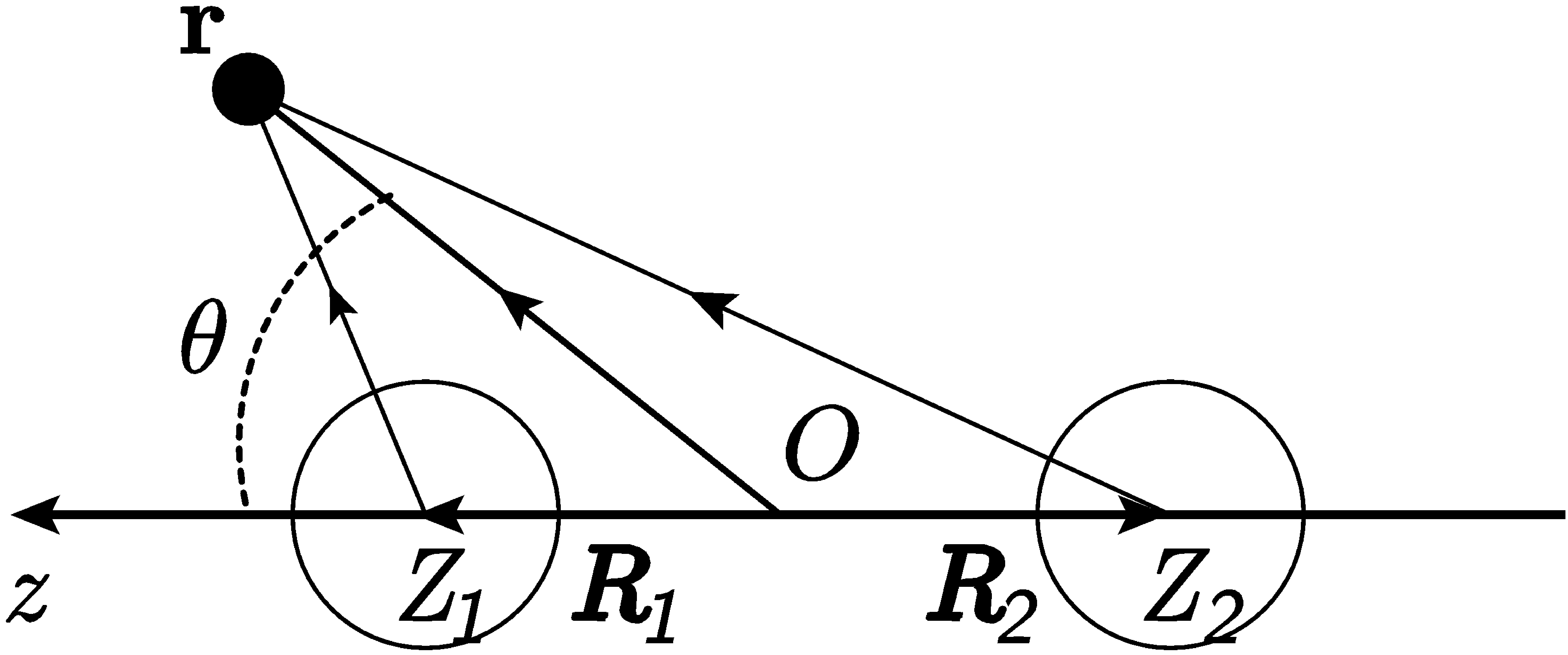}
\caption{The coordinate system for diatomic one-electron system oriented along the $z$-axis. The bound electron is marked with a filled circle, the nuclei are marked with circles with the nucleus charges $Z_1$ and $Z_2$. The corresponding position vectors from the coordinate origin $O$ are indicated as $\boldsymbol{r}$, $\boldsymbol{R}_1$ and $\boldsymbol{R}_2$. Since axial symmetry, the polar angle $\theta$ is introduced. }
\label{fig:2}
\end{figure}

To describe these simplest quasi-molecules, we can obtain the energy spectrum of the bound electron by solving the stationary Dirac equation in the Born-Oppenheimer approximation:
\begin{eqnarray}
\label{6}
\hat{H}_{\rm D}\psi_n= \varepsilon_n\psi_n,
\end{eqnarray}
where the Dirac Hamiltonian $\hat{H}_{\rm D}$ is
\begin{eqnarray}
\label{7}
\hat{H}_{\rm D} & = & \bm{\alpha}\cdot\mathbf{p} + V(Z_1,Z_2,\bm{r}) + \beta, \\
V(Z_1,Z_2,\bm{r}) & = & V(Z_1,\left|\bm{r}-\bm{R}_1\right|) +V(Z_2,\left|\bm{r}-\bm{R}_2\right|)
\end{eqnarray}
The binding potential $V$ is the sum of the two nuclear potentials corresponding to charge numbers $Z_1$ and $Z_2$. The electron energies $\varepsilon_n$ and wave functions $\psi_n$ depend parametrically on the internuclear distance $R$. The position vectors of the electron, $\bm{r}$, and of the nuclei, $\bm{R}_1$ and $\bm{R}_2$, are defined according to the coordinate system illustrated in Fig.~\ref{fig:2}.

We employ a method that has proven efficient for fully relativistic calculations of heavy quasi-molecules \cite{Kotov2020,Kotov2021,Kotov2022} and, more recently, for light one-electron systems \cite{Danilov_ADKB_2023,Solovyev_ADKB_2024}, including antiprotonic ones \cite{Anikin:23:jcp,Anikin_2025,Solovyev_ADKB_2025}. 
The dual kinetic balance (DKB) method for systems with axial symmetry (ADKB) is based on the decomposition of the required wave function over a finite basis set of B-splines \cite{Johnson_Bspline,Sapirstein_1996} with the imposed dual-kinetic-balance conditions \cite{Shabaev_DKB} generalized to the case of axially symmetric systems \cite{Rozenbaum_ADKB}.

The required bispinor in the spherical coordinate system can be written as
\begin{eqnarray}
\label{8}
\psi(r,\theta,\varphi) = \frac{1}{r}
\begin{pmatrix}
G_1(r,\theta) e^{\mathrm{i}(m_J-1/2)\varphi}
\\
G_2(r,\theta) e^{\mathrm{i}(m_J+1/2)\varphi}
\\
\mathrm{i}F_1(r,\theta) e^{\mathrm{i}(m_J-1/2)\varphi}
\\
\mathrm{i}F_2(r,\theta) e^{\mathrm{i}(m_J+1/2)\varphi}
\end{pmatrix},
\end{eqnarray}
where $m_J$ is the projection onto the molecular axis of the angular momentum operator $J$. The radial-polar components $G_i(r,\theta)$ and $F_i(r,\theta)$ ($i=1,2$) can be expanded in a finite basis set
\begin{eqnarray}
\label{9}
\phi(r,\theta)=
\begin{pmatrix}
G_1(r,\theta)
\\
G_2(r,\theta)
\\
F_1(r,\theta)
\\
F_2(r,\theta)
\end{pmatrix}
\\
\nonumber
\approx\sum\limits_{u=1}^4 
\sum\limits_{\substack{i_r=1 \\ i_\theta=1}}^{N_r, N_\theta} \Lambda C_{i_r i_\theta}^u B_{i_r}(r)Q_{i_\theta}(\theta)e_u,\qquad
\end{eqnarray}
where $B_{i_r}(r)$ and $Q_{i_\theta}$ are some basis functions, $e_u$ are the four-component basis unit vectors. The $\Lambda$ matrix imposes dual-kinetic-balance (DKB) relations between Dirac bispinor components, see \cite{Shabaev_DKB, Rozenbaum_ADKB} for details and the matrix form.

The expressions above are given for arbitrary basis sets $\{B_{i_r}(r)\}_{i_r=1}^{N_r}$ and $\{Q_{i_\theta}(r)\}_{i_\theta=1}^{N_\theta}$. The particular choice of one-component basis functions corresponds to $B$-splines as $B_{i_r}(r)$
and Legendre polynomials as $Q_{i_\theta}(\theta) = P_{i_\theta-1}\left(\frac{2\theta}{\pi}-1\right)$, forming the set of one-component $\theta$-dependent basis functions of polynomial degrees $l=0\dots N_\theta-1$. By varying the values $N_r$, $N_\theta$, such a sample allows the convergence of numerical calculations to be monitored. 

When the basis functions are constructed in this manner, the solution is sought inside a box, and the nucleus is treated as extended (we employ the Fermi model for the nuclear charge distribution). Consequently, the full Dirac spectrum of energies and eigenfunctions is replaced by a quasi-complete set consisting entirely of discrete states (including the negative part of the spectrum). The lowest of these states (whose number depends on the box size) agree, to within machine precision, with the analytical relativistic energies of the bound states.

It was previously shown in \cite{Solovyev_ADKB_2024} that the present ADKB approach makes it possible to reproduce the terms of light one-electron quasi-molecular compounds with sufficient accuracy (on the order of $10^{-8}$ in relative magnitude). In the present study, this fact is used to evaluate expression (\ref{4}) for the $H-p$ (hydrogen-proton) system.
Here, the state $a$ is chosen as the ground term, while the summation over $n$ is reduced to a summation over the pseudo-spectrum of discrete states.

\section{Numerical results}
\label{NR}

\subsection{Thermal potential curve and polarizability}
\label{h2+}

\begin{figure}[hbtp]
\centering
\includegraphics[width=0.7\columnwidth]{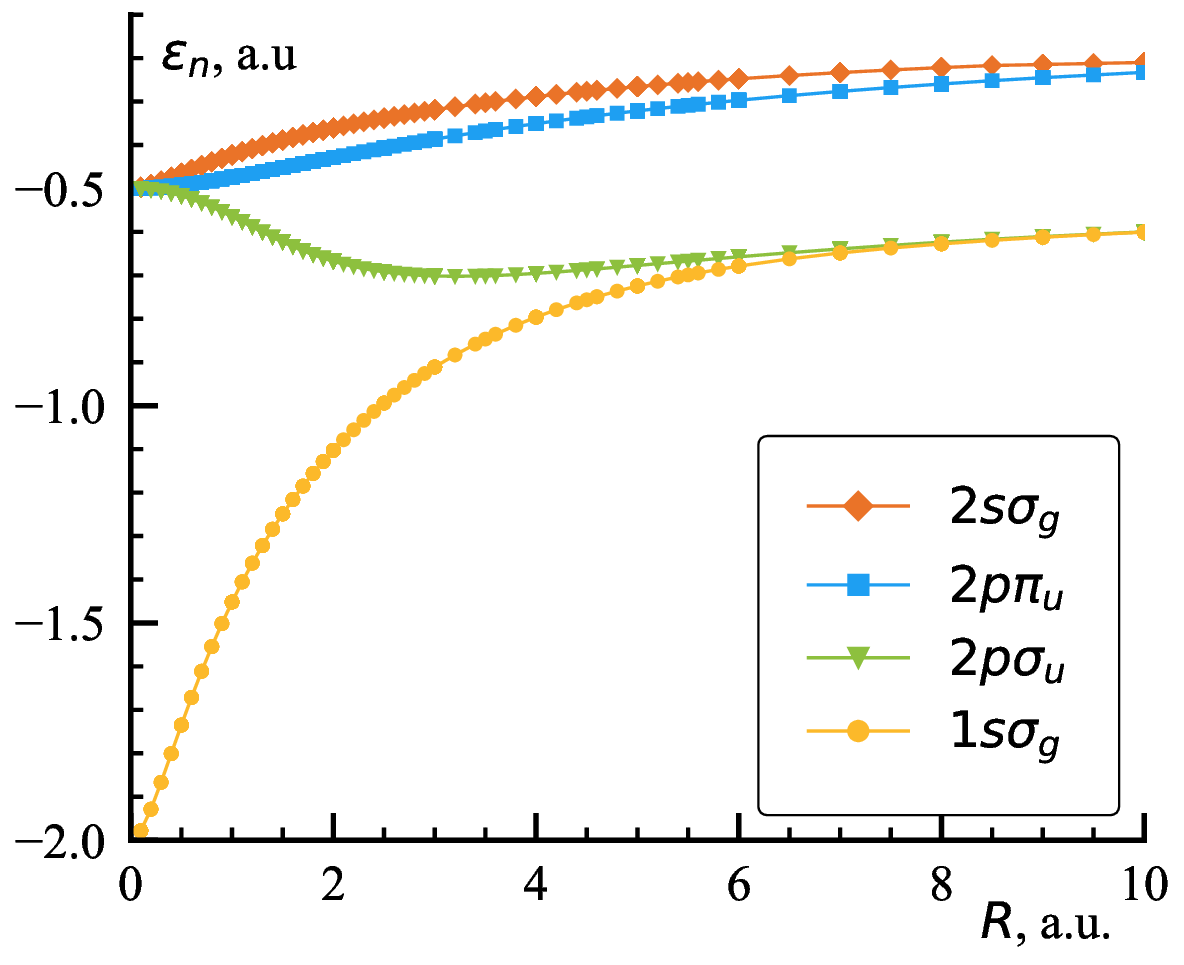}
\caption{The \textcolor{black}{relativistic} adiabatic potential curves of the first four lower states for the H-p quasi-molecular ion in dependence on inter-nuclear distance (abscissa axis). All values are given in atomic units.}
\label{fig:3}
\end{figure}

Recently, within the framework of the ADKB approach, the molecular hydrogen ion was studied in \cite{Solovyev_ADKB_2024}. In particular, it was demonstrated that the calculations performed enable the determination of molecular term energies 
with a relative accuracy of $2.08\times 10^{-8}$. In turn, this allows relativistic corrections to be extracted from the calculation with an accuracy ranging from $10^{-4}$ to $10^{-3}$. For brevity, here we present only a graphical representation of the \textcolor{black}{relativistic} ground term and the first three excited molecular terms of the H$_2^+$ system as a function of the internuclear distance, see Fig.~\ref{fig:3}. For a detailed analysis, refer to section III in \cite{Solovyev_ADKB_2024}.

We should note, however, two points. First, the presented graphs do not include the internuclear interaction (usually taken into account when obtaining potential energy curves). This interaction can nevertheless be easily incorporated in subsequent calculations by adding the corresponding $1/R$ terms, since $Z_1 = Z_2 = 1$. Second, the well-known splitting of the ground-state term into the $1s\sigma_g$ and $2p\sigma_u$ states occurs at about $R = 6$ atomic units. The latter becomes especially significant in calculations of dipole polarizabilities or the thermal shift in Eq.~(\ref{4}), because for the states $a = 1s\sigma_g$ and $n = 2p\sigma_u$ the energy denominator tends to zero as the internuclear distance approaches the "critical point" from zero. However, this term coalescence occurs sufficiently far from the equilibrium bond length of the ion (about $2$ atomic units). Hence, the calculated thermal shift values can be constrained by the corresponding potential well.

Numerical results for different inter-nuclear distances and temperature of $300$ K are collected in Table~\ref{tab:1} for the ground state of the $H-p$ ion. The number of digits is constrained by the precision of our numerical calculations and matches the accuracy of the vibrational level determination from the derived PECs (see below).
\begin{table}[H]
\renewcommand{\arraystretch}{1.1}
\caption{
Real part of the thermal shift from Eq.~(\ref{4}) at 300 K (in Hz) for the ground state of H$_2^+$ as a function of the internuclear distance $R$ (col.~1). Columns 2 and 3 list the contributions from $l=0$ and $l=1$ in Eq.~(\ref{5}); terms with $l\geqslant 2$ are of order $\lesssim 10^{-10}$ and omitted. \textcolor{black}{Column 4 gives the total self-energy contribution, column 5 lists the scalar polarizability $\alpha(0)$ in a.u., obtained as ADKB/dipole-approximation results, see Ref.~\cite{Lopez_2025} for theoretical details. 
In each row, the second line reports the results from the Ref.~\cite{Tsogbayar_2009} (the $\dag$-marked values), the energy shift is obtained as $\Delta \varepsilon_a^\beta =-\alpha_a(0) \langle E^2 \rangle / 2$, where the mean‑square field is
$\langle E^2\rangle=\frac{4 \pi^3}{15} \alpha^3 (k_{\rm B} T)^4\sim(831.9 \text{V/m})^2[\text{T}(\text{K})/300]^4$
\cite{Itano}.}}
\label{tab:1}
\centering
\begin{tabular}{l|l|l|l|l}
\hline
\hline
R & $\Delta_0 \varepsilon_a^\beta$, Hz & $\Delta_1 \varepsilon_a^\beta$, Hz & $\Delta \varepsilon_a^\beta$, Hz & $\alpha_a(0)$, a.u.\\
\hline
0.1 & -0.003772 & 0.001257 & -0.002515 & 0.29201633 \\
 & & & -0.00251495$^\dag$ &0.29205961$^\dag$ \\
\hline
0.2 & -0.004122 &	0.001372 & -0.002749 & 0.31882922 \\
 & & & -0.00274599$^\dag$ & 0.31889020$^\dag$ \\
\hline 
0.3 & -0.004613 & 0.001537 & -0.003075 & 0.35718718 \\
 &  &  &  -0.00307627$^\dag$ & 0.35724467$^\dag$ \\
 \hline 
0.4 & -0.005225 & 0.001744   & -0.003480 & 0.40523962 \\
 &  &  &  -0.00349004$^\dag$ & 0.40529533$^\dag$ \\
 \hline
0.5 & -0.005974 & 0.001990 & -0.003983 & 0.46233361 \\
 &  &  & -0.00398167$^\dag$ & 0.46238807$^\dag$ \\
 \hline
0.6 & -0.006820 & 0.002275 & -0.004545 & 0.52839725 \\
 &  & & -0.00455055$^\dag$ & 0.52845176$^\dag$ \\
 \hline
0.7 & -0.007748 & 0.002599 & -0.005149 & 0.60369360 \\
 &  &  & -0.00519893$^\dag$ & 0.60374805$^\dag$ \\
 \hline
0.8 & -0.008625 & 0.002965 & -0.005659 & 0.68870223 \\
 & & & -0.00593097$^\dag$ & 0.68875929$^\dag$ \\
 \hline
0.9 & -0.009894 & 0.003376 & -0.006518 & 0.78407541 \\
 &  &  & -0.00675229$^\dag$ & 0.78413766$^\dag$ \\
 \hline
1.0 & -0.01130 & 0.00383 & -0.00747 & 0.89061503 \\
 &  &  & -0.00766977$^\dag$ & 0.89068427$^\dag$ \\
\hline 
1.3 & -0.01649 & 0.00554 & -0.01095 & 1.2874314 \\
 &  &  & -0.0110871$^\dag$ & 1.2875344$^\dag$ \\
\hline
1.5 & -0.02093 & 0.00701 & -0.01391 & 1.6293207 \\
 &  &  & -0.0140314$^\dag$ & 1.6294607$^\dag$ \\
\hline
1.8 & -0.02953 & 0.00987 & -0.01965 & 2.2930753 \\
 &  & & -0.0197479$^\dag$ & 2.2933028$^\dag$ \\
\hline
2.0 & -0.03700 & 0.012332 & -0.02467 & 2.8639987 \\
 &  &  & -0.0246649$^\dag$ & 2.8643154$^\dag$ \\
\hline
3.0 & -0.11000 & 0.03670 & -0.07329 & 8.5225662 \\
 &  &  & -0.0734031$^\dag$ & 8.5242412$^\dag$ \\
\hline
4.0 & -0.336 & 0.11204 & -0.223 & 25.992918 \\
&  &  & -0.2238198$^\dag$ & 25.992018$^\dag$ \\
\hline
5.0 & -1.069 & 0.35640 & -0.712 & 82.211317 \\
&  &  & -0.7078922$^\dag$ & 82.206974$^\dag$ \\
\hline
6.0 & -3.51 & 1.1767 & -2.33 & 262.90865 \\
 &  & & -2.2637803$^\dag$ & 262.89104$^\dag$ \\
\hline
7.0 & -14.09 & 4.69 & -9.40 & 832.86184 \\
&  &  & -7.1713064$^\dag$ & 832.79821$^\dag$ \\
\hline
8.0 & -25.175 & 8.396 & -16.779 & 2596.28908\\
&  &  & -22.355216$^\dag$ & 2596.0938$^\dag$ \\
\hline
9.0 & 20.8 & -6.9 & 13.9 & 7968.9629 \\
&  &  & -68.618819$^\dag$ & 7968.6499$^\dag$ \\
\hline
10.0 & 29.96 & -9.991 & 19.97 & 24140.871 \\
&  &  & -207.90248$^\dag$ & 24143.553$^\dag$ \\
\hline
\hline
\end{tabular}
\end{table}


The numerical values reported in Table~\ref{tab:1} retain a stable number of digits under the following parameter variations. First, convergence is assessed by changing $N_r$ and $N_\theta$; accordingly, the number of basis functions was set to yield six significant digits after the decimal point for short distances. Second, the box size was chosen to secure eight significant digits in the hydrogen atom energies at the limit of infinite internuclear separation (these energies were checked against the well-known Sommerfeld formula for hydrogen). As a benchmark, computations were carried out for two nuclear models (the Fermi and the shell model) and the results agreed to within $10^{-11}$ in relative value for \textcolor{black}{$R = 0.1,~1$} and $10$ a.u. Third, the frequency integral in Eq.~(\ref{4}) was evaluated while varying the number of Gaussian quadrature points, which ensured the stability of all quoted digits. For the summation over intermediate states, $n = 120$ levels from the obtained pseudospectrum were used. \textcolor{black}{The value of $n$ was chosen from a convergence study. Since the results are stable for this basis set size, we regard it as sufficient to ensure convergence.}

A rapid convergence of the partial-wave expansion is observed first when computing the thermal shift of the ground state. This behavior is explained by the dipole approximation, which proves highly appropriate here \cite{Lopez_2025}. For such a system, the main contribution comes from opposite-parity matrix elements. Specifically, Table~\ref{tab:1} shows that the monopole ($l=0$) and dipole ($l=1$) contributions are comparable in magnitude but opposite in sign for the partial-wave expansion (\ref{5}). It is precisely their cancellation that yields the appropriate result for the thermal shift. We, however, performed the calculation of the quadrupole moment, $l=2$, and found its value to be negligible at all inter-nuclear distances. 

The dipole approximation within the ADKB approach also serves to verify the scalar polarizability values $\alpha_a(0) = \frac{2e^2}{3}\sum_{n\neq a}\frac{|\langle n|\bm{r}|a\rangle|^2}{E_n-E_a}$ (see the last column of Table~\ref{tab:1}). The latter allows one to determine the thermal Stark shift via the known relation $\Delta E = -\frac{1}{2}\alpha_a(0)\bm{E}^2$ using the mean-square field of blackbody radiation, $\langle E^2\rangle=(831.9 \text{ V/m})^2[\text{T}(\text{K})/300]^4$ \cite{farley,Itano}. \textcolor{black}{Using this relation, we convert the polarizability data from Ref.~\cite{Tsogbayar_2009} into thermal Stark shifts and compare them with the self-energy correction obtained in this work. The corresponding results are listed in the fourth column.} 

\begin{figure}[hbtp]
\centering
\includegraphics[width=0.75\columnwidth]{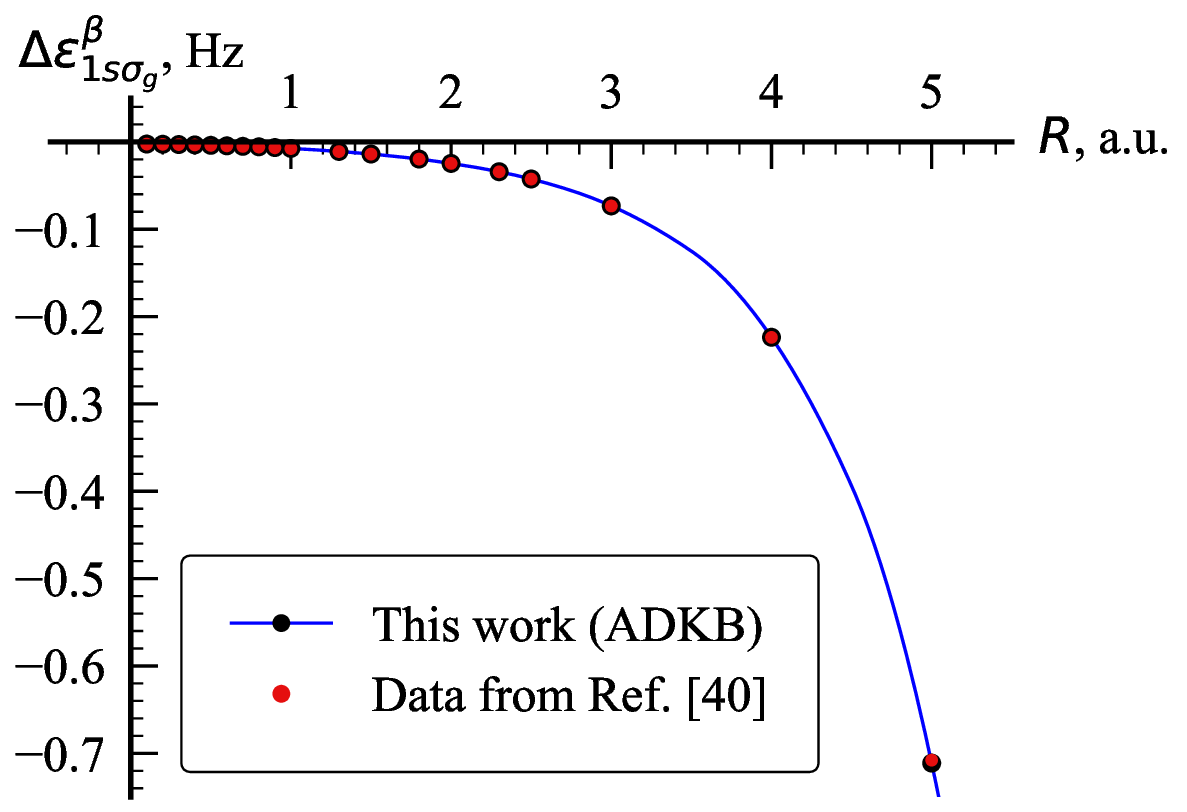}
\caption{Thermal shift (\ref{4}) as a function of internuclear distance, constructed from the values presented in Table~\ref{tab:1}. The solid curve passing through the points (table values) is obtained using second‑order spline interpolation. The abscissa is the internuclear distance, and the ordinate is the thermal shift. The thermal shift values are given in Hz, whereas the internuclear distance is specified in atomic units.}
\label{fig:4}
\end{figure}

\begin{figure}[hbtp]
\centering
\includegraphics[width=0.7\columnwidth]{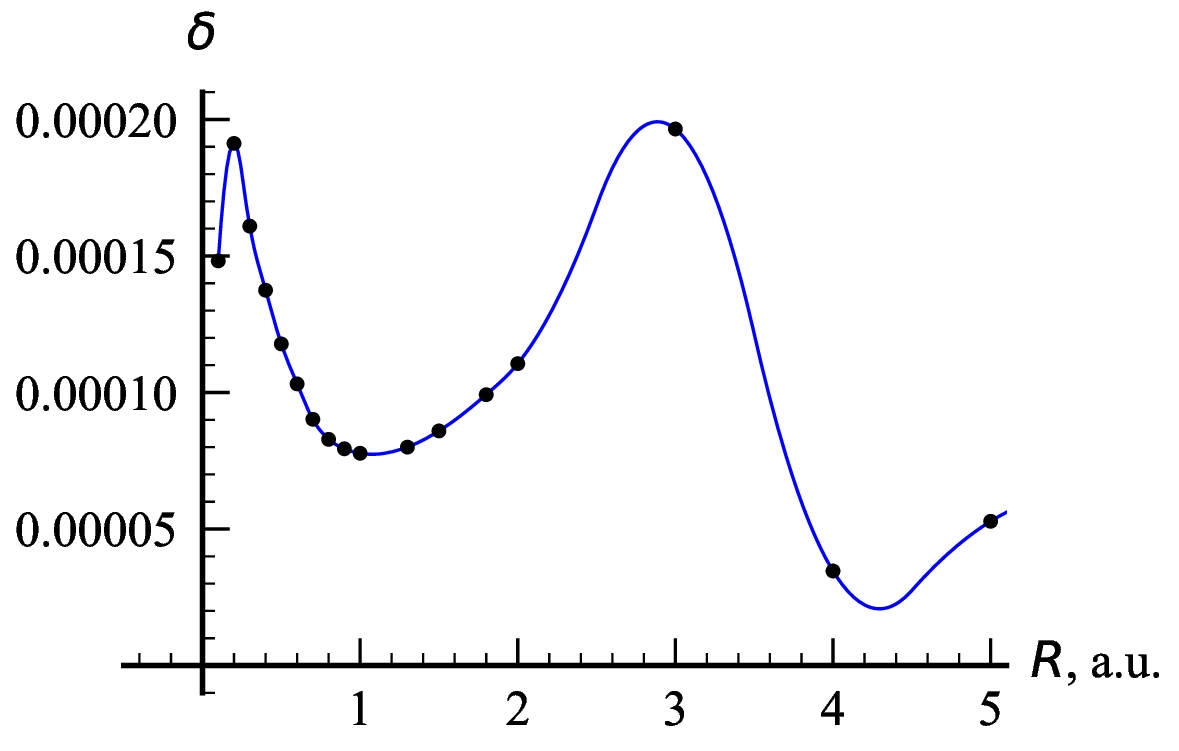}
\caption{Graph showing the relative deviation of the scalar dipole polarizability calculated within the ADKB approach from the values reported in \cite{Tsogbayar_2009}.\textcolor{black}{The data in this figure correspond to the last column of Table~\ref{tab:1}}}
\label{fig:6}
\end{figure}

Although \textcolor{black}{the total thermal self-energy correction} agree to within about one percent for the thermal shift in the vicinity of the potential well, \textcolor{black}{see Fig.~\ref{fig:4}}, the calculated \textcolor{black}{shifts} differ markedly from those reported in \cite{Tsogbayar_2009}. \textcolor{black}{This is especially visible} for $R \geqslant 6$ a.u., where they even exhibit the opposite sign at large internuclear distances. For instance, the relative difference amounts to approximately $0.7\%$ at $R=5$ a.u., $3.8\%$ at $R=6$ a.u., and rises above $25\%$ for $R\geqslant 8$ a.u. A clear increase of the \textcolor{black}{total thermal Stark shift} with growing internuclear distance is observed. This behavior is directly evident from Fig.~\ref{fig:3}, which unambiguously shows the vanishing energy difference between the $1s\sigma_g$ and $2p\sigma_u$ states starting at separations of about six atomic units. \textcolor{black}{In turn, relative difference for the scalar dipole polarizability $\alpha_a(0)$ values does not exceed $~2\times10^{-4}$ in the $R = (0.1; 10)$ a.u. region, see Fig.~\ref{fig:6}.}

Despite this discrepancy with Ref.~\cite{Tsogbayar_2009}, our results from the two approaches (partial-wave decomposition and the direct dipole-approximation calculation) are consistent for all distances, including $R>6$ a.u. However, the calculation of the thermal shift within the static limit and dipole approximation is not strictly justified, since the vanishing energy splitting between the lowest states effectively reduces the power of the frequency prefactor in Eq.~(\ref{4}). Nevertheless, the thermal shifts computed using this approach are reliable up to at least $R = 6$ a.u., the most relevant region of the H$_2^+$ potential well. The distance dependence and the tabulated data point to the necessity of accurately accounting for frequency dependence. The dynamic nature of the thermal shift is thus of central importance.

Note that as $R\rightarrow\infty$, the scalar dipole polarizability should reduce to the hydrogen atomic value of $9/2$, although this asymptotic limit is not strictly enforced in our numerical calculation. The thermal shift of the hydrogen ground state is likewise well known, equal to $-0.03875176$ Hz \cite{Reiter_2026}. The data presented in Table~\ref{tab:1} and the nonrelativistic calculation of Ref.~\cite{Tsogbayar_2009} do not support these two limiting values. Nonetheless, we note that for the calculation of thermal shifts of vibrational levels (i.e., energy levels within the potential well of the hydrogen molecular ion), the results show satisfactory consistency. To clarify the limiting behavior, we have performed fully relativistic calculations for larger internuclear separations. The obtained values are not included in Table~\ref{tab:1} owing to the reduced convergence of numerical calculations at large internuclear distances within the ADKB approach. Nevertheless, we are confident that they exhibit the correct qualitative behavior, and the results are therefore presented graphically below.

\begin{figure}[hbtp]
\centering
\includegraphics[width=0.7\columnwidth]{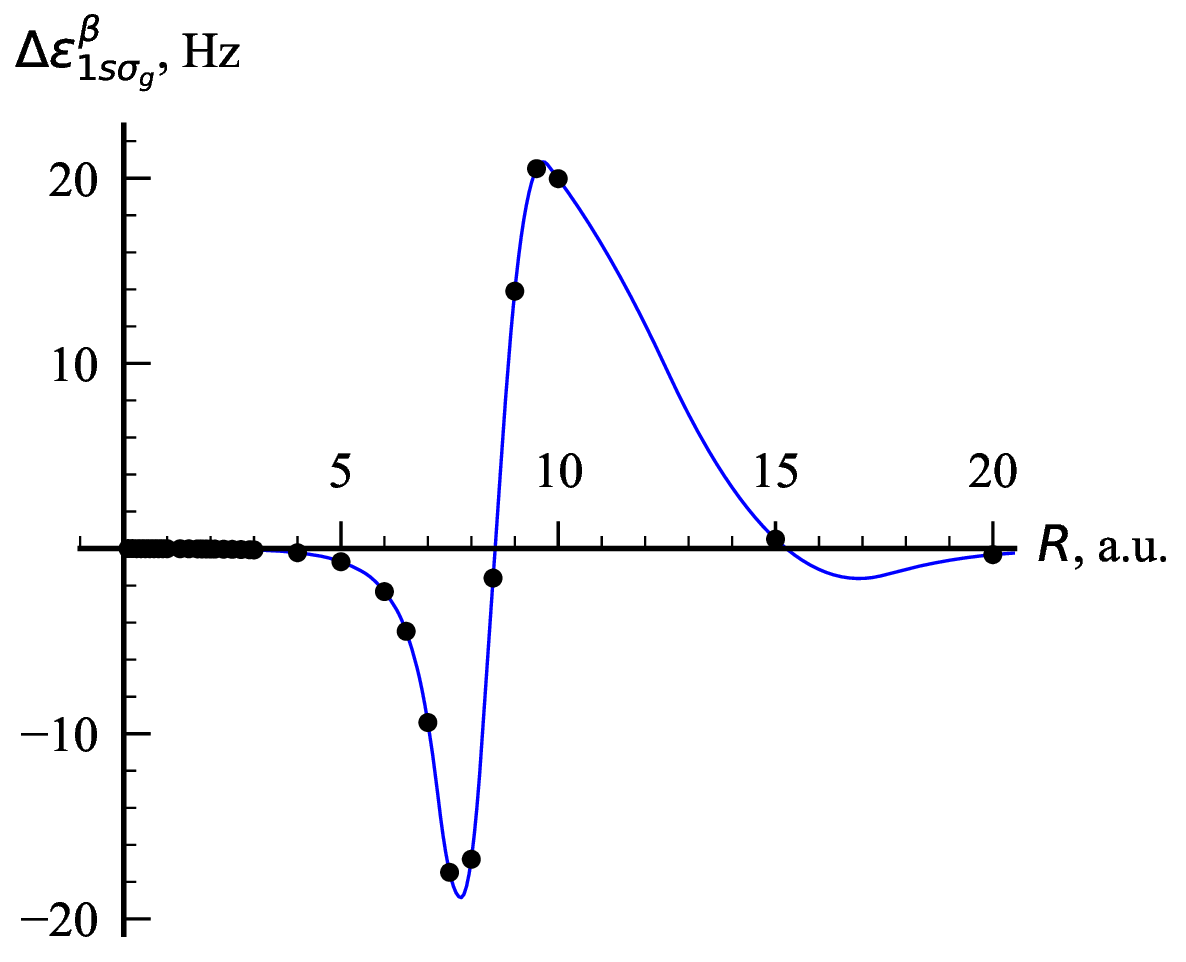}
\caption{The thermal shift is calculated exactly using Eq.~(\ref{4}), which accounts for the dynamical nature of the effect. The numerical results are shown as a function of the internuclear distance. The shift values are given in Hz, and the distance is in atomic units.}
\label{fig:5}
\end{figure}

The thermal shift of the ground term as a function of the internuclear distance is shown in Fig.~\ref{fig:4} for the range $R\in [0,5]$ a.u. By performing calculations for larger $R$, one can observe that $\Delta \varepsilon_{1s\sigma_g}^\beta$ asymptotically approaches the negative value $-0.03875176$ Hz, see Fig.~\ref{fig:5}. 

We defer more accurate calculations for larger distances, as they would require additional software development (Fig.~\ref{fig:5} provides the qualitative behavior). Instead, we focus on a detailed description of the effect in the interatomic bonding region. To compare our results with the literature, we list in the last column of Table~\ref{tab:1} the scalar polarizabilities computed using the ADKB approach in the dipole approximation, alongside the values taken from Ref.~\cite{Tsogbayar_2009}. 

The relative absolute difference of the static dipole polarizabilities, $\delta=|(\alpha_a^{\rm ADKB}(0)-\alpha_a^{\text{\cite{Tsogbayar_2009}}}(0))/\alpha_a^{\text{\cite{Tsogbayar_2009}}}(0)|$, is shown in Fig.~\ref{fig:6} to illustrate the discrepancies between the two approaches.
As shown in the figure, the relative difference between the static dipole polarizabilities obtained from the ADKB approach and those reported in \cite{Tsogbayar_2009} fluctuates within $0.01\%$ for $R\in[0,6]$ a.u. Hence, the ADKB numerical calculation is essentially identical to the well-established nonrelativistic computational method~\cite{Korobov_2000}. The discrepancy in the potential-well region is negligible for our purposes. For internuclear distances $R \gtrsim 8$ a.u., the behavior of $\Delta\varepsilon_a^\beta$ is determined exclusively by the dynamical character of the shift. The deviation becomes significant, with the exponential dependence on $R$ being replaced by damped oscillations. This behavior is explained by the degeneracy of the $1s\sigma_g$ and $2p\sigma_u$ molecular terms, see Fig.~\ref{fig:3}.

\begin{figure}[hbtp]
\centering
\includegraphics[width=0.7\columnwidth]{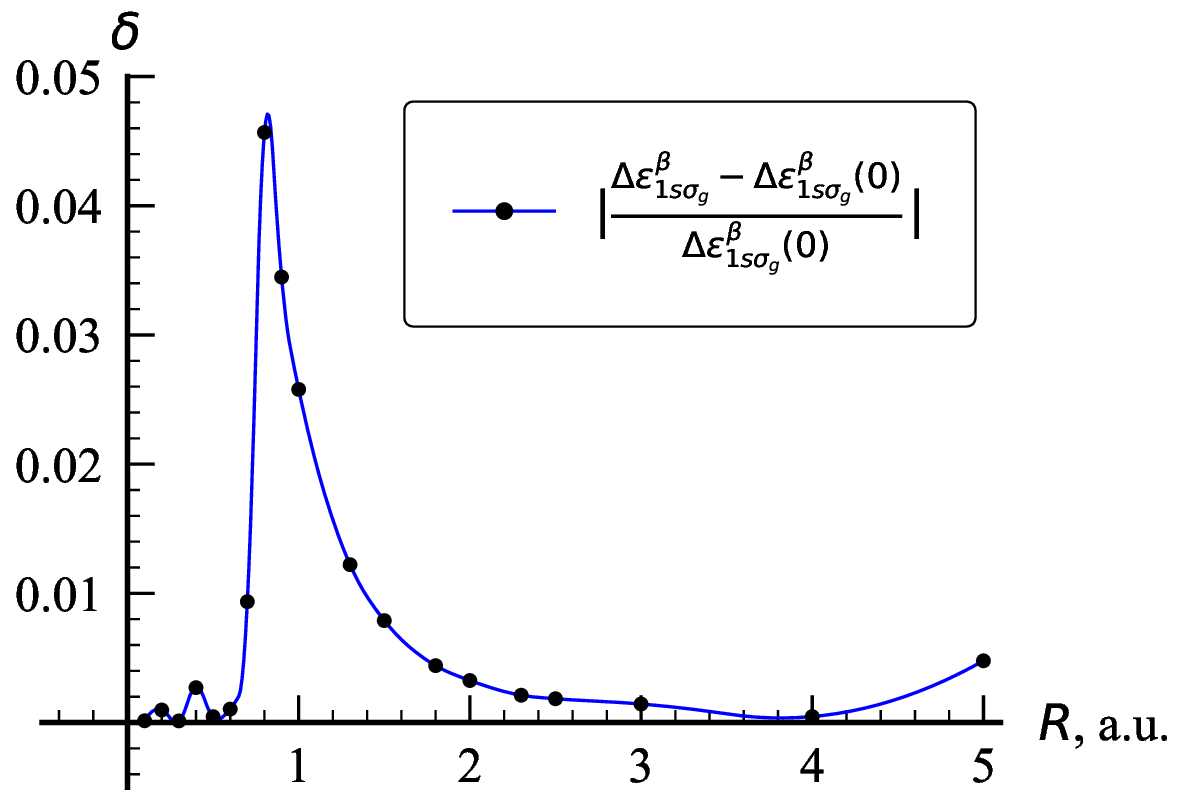}
\caption{Relative difference of the thermal shift calculated from the full dynamic expression and its static limit. \textcolor{black}{The data in this figure correspond to the fourth column of Table~\ref{tab:1}}}
\label{fig:7}
\end{figure}

This analysis was devoted to the calculation and comparison of scalar polarizabilities within the dipole approximation, which is commonly used for determining thermal shifts. However, as recently demonstrated in \cite{Reiter_2026,Bobylev_2026}, careful frequency integration and the inclusion of higher multipoles, see, e.g., \cite{Lopez_2025} and references therein, can play a crucial role. The dynamic nature of the thermal shifts, as expressed by Eq.~(\ref{4}), can lead to deviations from the static-limit results that substantially exceed the dynamic corrections themselves. As an illustration, Fig.~\ref{fig:7} shows the relative deviation between the thermal shift obtained from the full frequency-dependent calculation using Eq.~(\ref{4}) and that obtained in the static limit (see Table~\ref{tab:1}).
As follows from the figure, the deviation of the dynamic shift from its static-limit value amounts to several percent at the edges of the potential well, which is considerably larger than the dynamic corrections, see \cite{Schiller_P}. One may therefore expect a comparable change in the thermal shifts of the vibrational levels, of the order of one percent. A detailed calculation of these shifts for the hydrogen molecular ion is given in the following section.

\begin{figure}[hbtp]
\centering
\includegraphics[width=0.8\columnwidth]{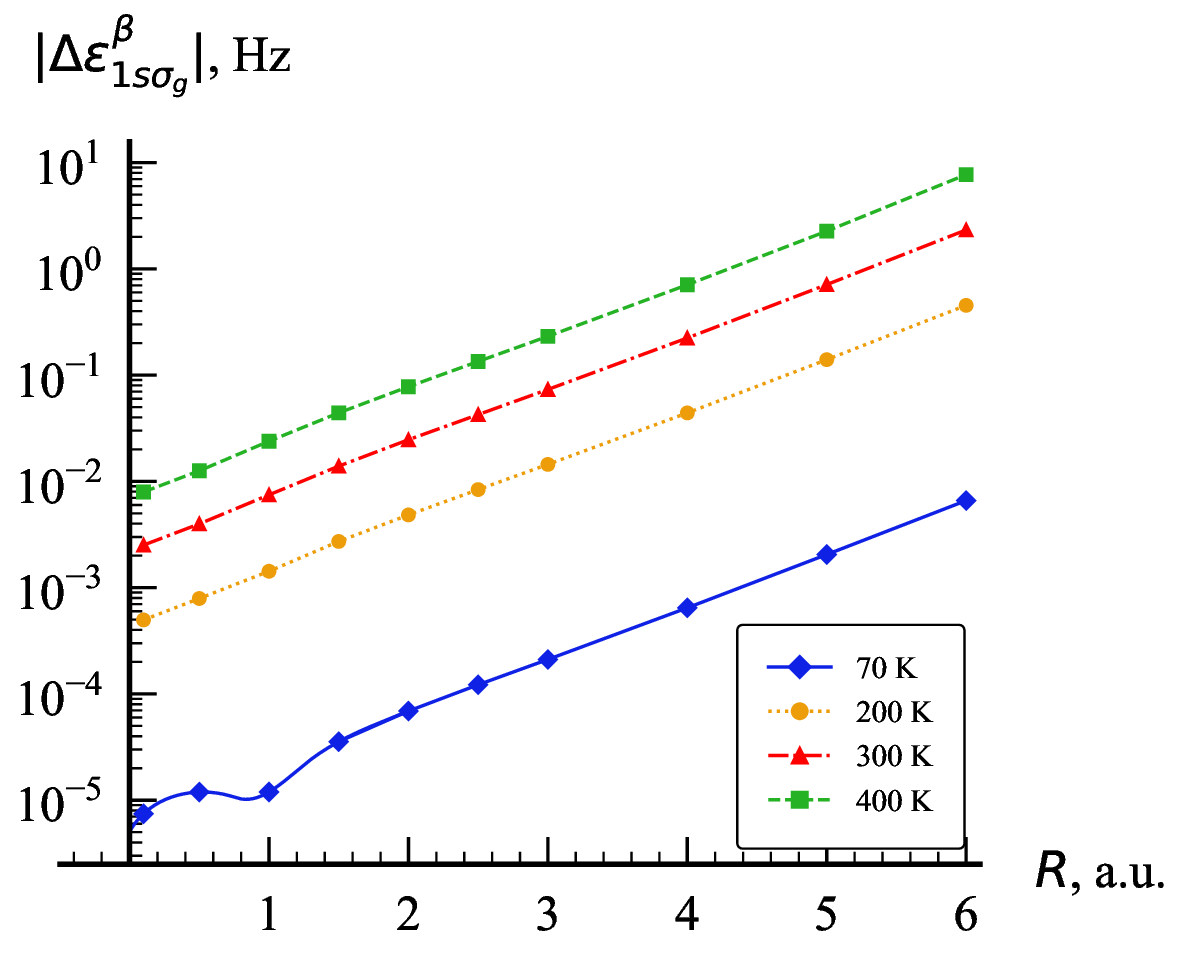}
\caption{The graph displays the thermal shift of Eq.~(\ref{4}) as a function of the internuclear distance for several temperatures. The absolute value of the shift is shown on a logarithmic scale.}
\label{fig:8}
\end{figure}

On the basis of the above analysis, we present below the thermal shifts calculated for the ground term of the H$_2^+$ ion at different temperatures, see Table~\ref{tab:2}.
\begin{table}[H]
\caption{The table lists the thermal shift as a function of the internuclear distance $R$ for the $1s\sigma_g$ state of the H$_2^+$ molecular ion at different temperatures. The first column shows the internuclear distance $R$ in atomic units. The second, third, and fourth columns give the energy shift from Eq.~(\ref{4}) in Hz, with the corresponding temperature indicated in parentheses.}
\label{tab:2}
\centering
\begin{tabular}{l|c|c|c}
\hline
$R$, a.u. & $\Delta \varepsilon_a^\beta$ ($70$ K), Hz & $\Delta \varepsilon_a^\beta$ ($200$ K), Hz & $\Delta \varepsilon_a^\beta$ ($400$ K), Hz  \\
\hline
\hline
0.1 & -0.000007	& -0.000497 & -0.007946 \\
0.5 & -0.000012 & -0.000791 & -0.012587 \\
1.0 & -0.000012 & -0.001427 & -0.023887 \\
1.5 & -0.000036 & -0.002722 & -0.044147 \\
2.0 & -0.000069 & -0.004838	& -0.077819 \\
2.5 & -0.000122 & -0.008375 & -0.134459 \\
3.0 & -0.000211 & -0.014447 & -0.231868 \\
4.0 & -0.000646 & -0.044087 & -0.708499 \\
5.0 & -0.002051 & -0.139832 & -2.264984 \\
6.0 & -0.006602 & -0.453511 & -7.711249 \\
\hline 
\hline
\end{tabular}
\end{table}
In addition to the table, the results are also presented graphically in Fig.~\ref{fig:8}.
In particular, from the data in Table~\ref{tab:2} and the plots in Fig.~\ref{fig:8}, it follows that the $(k_{\rm B}T)^4$ dependence actually serves as a scaling factor in the range of internuclear distances covering the potential well of the hydrogen molecular ion. Plotting the absolute values of the shift on a logarithmic scale at different temperatures also makes it clear that one should hardly expect "magic" temperatures at which the thermal shift of transition frequencies would vanish. A change in temperature merely results in a parallel shift of the curve, without altering its shape \textcolor{black}{over a wide temperature range}, unlike the atomic-physics case \cite{PhysRevA.100.023417}.

\subsection{Thermal shift of the vibrational spectrum}
\label{VTS}

Several methods can be used to calculate the thermal shift of vibrational levels. The first approach is to insert the computed $R$-dependent thermal shift directly into the Schr\"odinger equation that contains the nuclear kinetic energy, the internuclear Coulomb potential, and the adiabatic potential, see, e.g., Fig.~\ref{fig:3} for the latter. Nevertheless, since the shift is extremely small ($\sim 10^{-18}$ a.u.), such a procedure demands special numerical treatment, including at least double-precision computations. We proceed along a different route. 

First, we begin by determining the vibrational energies and their corresponding wave functions. As input, we use the commonly accepted nonrelativistic adiabatic potentials for the $1s\sigma_g$ state~\cite{Korobov_2006,Korobov_2007,Fernandez_2021}. In principle, these potentials can be supplemented by including relativistic and QED corrections~\cite{Korobov_2006,Korobov_2007}. It should be noted that the variational nonrelativistic calculations accurately account for the nonadiabatic motion of the nuclei and their finite masses. However, according to perturbation theory, such corrections are expected to affect the wave functions and energies only at the level of the fifth or sixth significant digit. Hence, this may impose the main limitation on the accuracy of our final thermal-shift values for the vibrational levels. For the eigenstates obtained from the one-dimensional vibrational problem, the thermal shift is given by the expectation value of $\Delta \varepsilon_a^\beta(R)$ for each vibrational level.

To calculate vibrational states in one-electron diatomic systems, we have developed a program that solves the vibrational problem within the ADKB method. Our calculations also account for nuclear motion effects through the diagonal Born-Oppenheimer correction (DBOC). This correction was evaluated using an extended implementation of the general ADKB approach via a dedicated code (finite-difference method). Upon explicit verification, it was found that this approach yields DBOC values with an accuracy within a few tenths of a percent. We then obtain the eigenvalues of the radial Schr\"odinger equation within the ADKB approach. The resulting energy spectrum is compared with the data produced by the DUO program \cite{Duo}, for which we adopt the commonly accepted values from Ref.~\cite{Korobov_2007}. \textcolor{black}{Thus, we compare the fully relativistic electronic ADKB results, which include the DBOC used to solve the nuclear Schr\"odinger equation, with independent results from the DUO program, using the energies from Ref.~\cite{Korobov_2007} as input. We note that the latter do not account for nuclear motion corrections and include relativistic corrections up to order $m\alpha^6$.} The results of calculation are listed in Table~\ref{tab:3}.
\begin{table}[ht!]
\caption{The table lists the vibrational energy levels of the H$_2^+$ ion for the $1s\sigma_g$ term. The first column shows the vibrational quantum number. The second and third columns present the ADKB energies without and with the diagonal Born-Oppenheimer correction, respectively. The fourth column provides the DUO results for comparison and last column contains relative deviation for the two approaches \textcolor{black}{(DUO and ADKB)}. The energies are shifted so that the $\upsilon=0$ level is at zero energy. All values are given in cm$^{-1}$ and rounded to two decimal places. Proton to electron mass ratio used in the ADKB is $1836.152\, 673\, 426$, see \cite{RevModPhys.97.025002}. Total angular momentum and it's projection onto molecular axis were taken as $1/2$.}
\label{tab:3}
\renewcommand{\arraystretch}{1.35}
\begin{tabular}{r|c|c|c|c} 
\hline
\hline
$\upsilon$ & $\varepsilon_\upsilon$, cm$^{-1}$ & $\varepsilon^{\rm DBOC}_\upsilon$, cm$^{-1}$ & $\varepsilon^{\rm DUO}$, cm$^{-1}$ & $\delta$ in $\%$ \\
\hline
0 & 0.0 & 0.0 & 0.0 & --\\
1 & 2191.68 & 2190.95 & 2191.46 & -0.0233 \\
2 & 4256.25 & 4254.90 & 4255.65 & -0.0176 \\
3 & 6198.92 & 6197.05 & 6196.80 & +0.0040 \\
4 & 8023.89 & 8021.60 & 8018.48 & +0.0389 \\
5 & 9730.55 & 9727.93 & 9723.59 & +0.0446 \\
6 & 11319.51 & 11316.69 & 11314.44 & +0.0199 \\
7 & 12795.59 & 12792.66 & 12792.68 & -0.0002 \\
8 & 14161.95 & 14159.02 & 14159.31 & -0.0021 \\
9 & 15417.15  & 15414.30 & 15414.66 & -0.0023 \\
10 & 16560.14 & 16557.48 & 16558.34 & -0.0052 \\
11 & 17590.83 & 17588.47 & 17589.21 & -0.0042 \\
12 & 18506.71 & 18504.74 & 18505.26 & -0.0028 \\
13 & 19303.89 & 19302.42 & 19303.64 & -0.0063 \\
14 & 19979.83 & 19978.99 & 19980.35 & -0.0068 \\
15 & 20528.02 & 20527.91 & 20529.92 & -0.0098 \\
16 & 20942.91 & 20943.66 & 20946.27 & -0.0125 \\
17 & 21218.34 & 21220.03 & 21221.98 & -0.0092 \\
18 & 21254.06 & 21257.35 & 21361.21 & -0.4862 \\
19 & 21309.09 & 21312.38 & 21401.10 & -0.4146 \\
\hline
\hline
\end{tabular}
\end{table}

Table~\ref{tab:3} first reveals that our method and the ADKB approach yield vibrational-level values that are in excellent agreement with the DUO reference data for low-lying levels, while significant differences appear for $\upsilon \geqslant 18$. The relative difference, given in percent in the last column of the table, does not exceed $0.04\%$ for $\upsilon \leqslant 17$. It can also be worth noting the good agreement of our values for the first three vibrational levels with the results reported in \cite{Pilon_rovib}. One can also compare our energy differences with those of \cite{Ishikawa_2012}. Specifically, for the $0-1$, $1-2$, and $2-3$ transitions from Table~\ref{tab:3}, the relative deviations from the reference data are about $0.036\%, 0.205\%, 0.795\%$, respectively. It should be noted, however, that the results in Table 3 of \cite{Ishikawa_2012} lack the diagonal Born-Oppenheimer corrections, which are important for these transition frequencies. Furthermore, the fundamental $0\to1$ transition frequency, $2190.95$ cm$^{-1}$, is consistent with the value $65688323710.1(5)(2.9)$ kHz reported in \cite{PhysRevLett.118.233001} within $0.008\%$. Thus, we conclude that our computed expectation values of the vibrational energies are determined with an accuracy of about $0.04\%$ at the present stage.

\begin{figure}[hbtp]
\centering
\includegraphics[width=0.8\columnwidth]{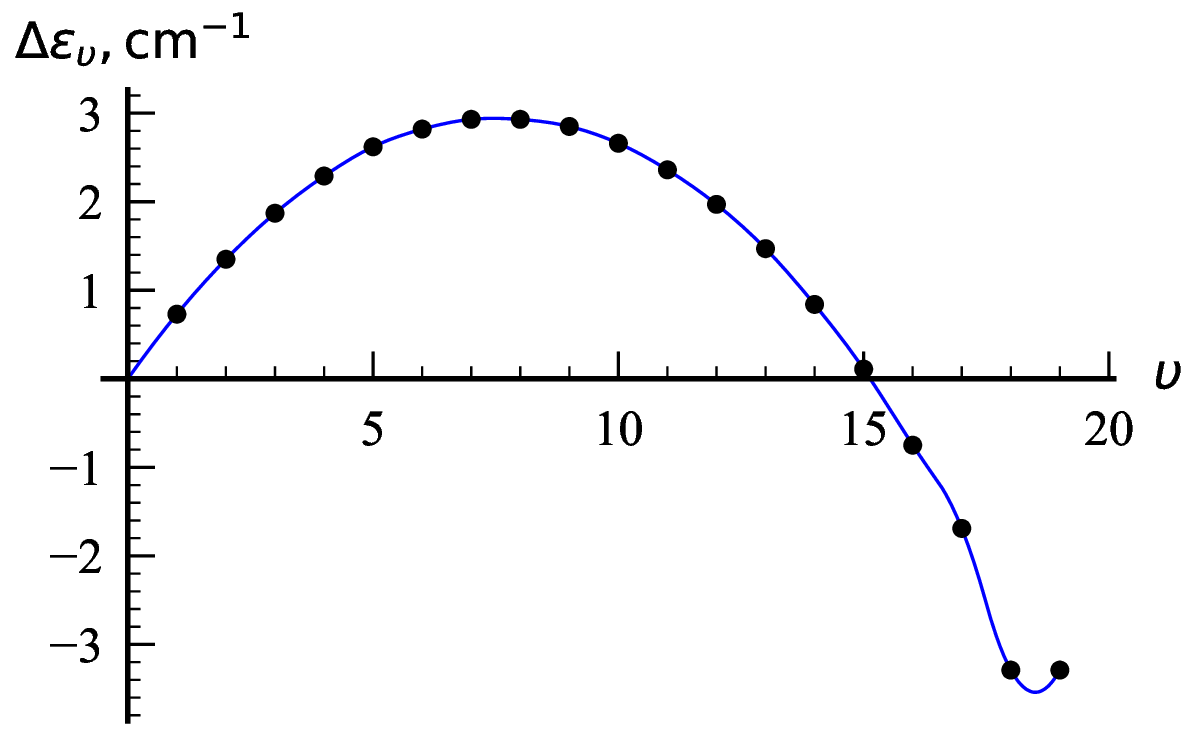}
\caption{Diagonal Born-Oppenheimer correction $\Delta\varepsilon_{\upsilon}$ vs the vibrational quantum number, obtained as the difference between columns 2 and 3 of Table~\ref{tab:3} (in cm$^{-1}$).}
\label{fig:9}
\end{figure}

The diagonal Born-Oppenheimer correction to the vibrational levels is shown in Fig.~\ref{fig:9} as a function of the vibrational quantum number $\upsilon$. This shift is obtained as the difference between the values in the second and third columns \textcolor{black}{of Table~\ref{tab:3}}, with all results compared within the ADKB framework.
In particular, the graph exhibits the characteristic DBOC maximum for homonuclear systems. Accordingly, the diagonal Born-Oppenheimer correction due to nuclear motion is substantially diminished in transitions between the corresponding vibrational levels. This $\upsilon$-dependence originates from the DBOC's variation with $R$. The DBOC as a function of $R$ has a minimum, displaced from the minimum of the potential well (the sum of the adiabatic electronic potential and the internuclear repulsion), resulting in different values of the correction for different $\upsilon$.

\begin{figure}[hbtp]
\centering
\includegraphics[width=1.0\columnwidth]{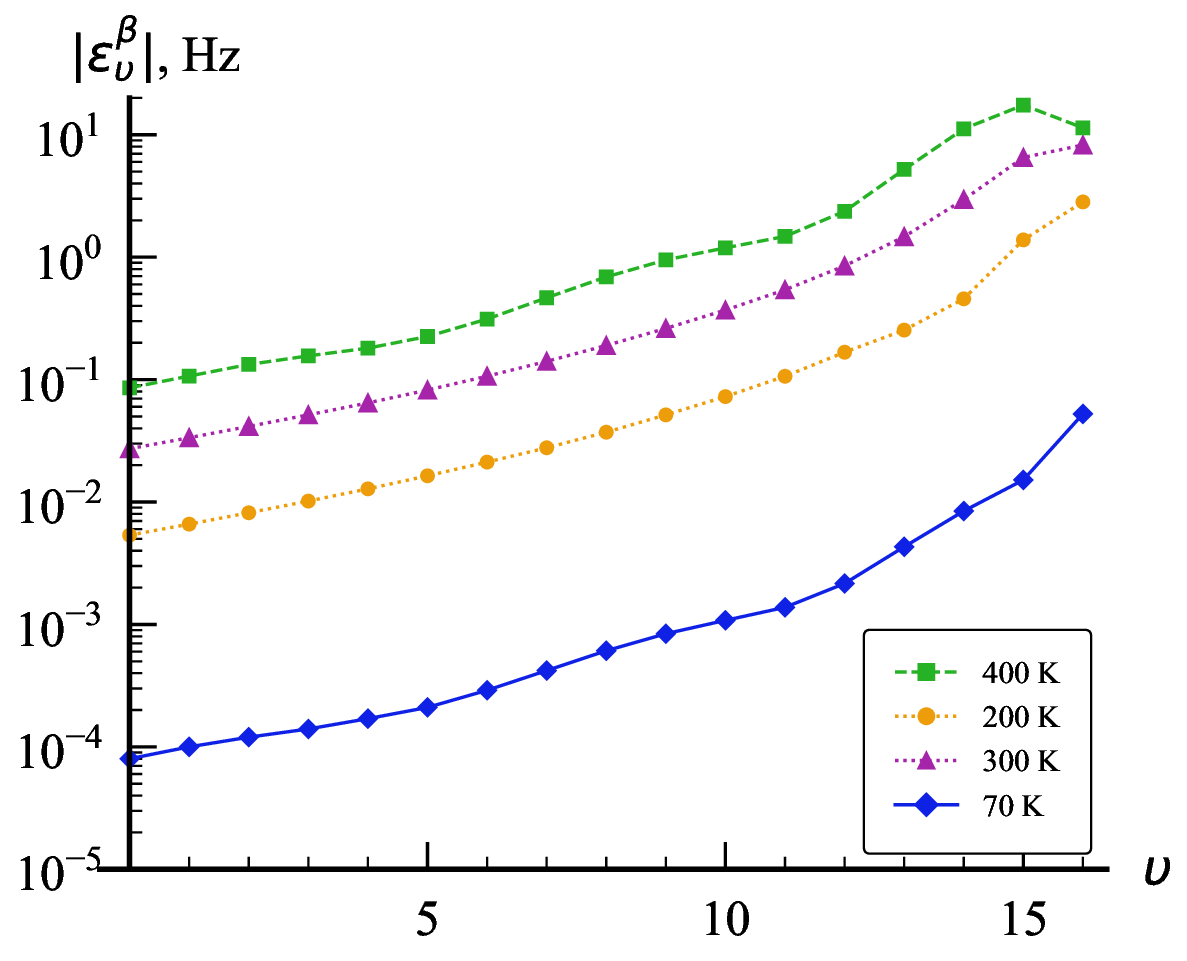}
\caption{The graph shows the expectation values of the thermal shift from Eq.~(\ref{4}) as a function of the vibrational quantum number at temperatures $70$, $200$, $300$, and $400$ K. The absolute value is plotted on a logarithmic scale.}
\label{fig:10}
\end{figure}

The expectation values of the thermal shift from Eq.~(\ref{4}), denoted as $\varepsilon_\upsilon^\beta$, are listed in Table~\ref{tab:4} for the set of vibrational levels from Table~\ref{tab:3}. The same table also gives the thermal shift values in Hz calculated at different temperatures.
\begin{table}[ht!]
\caption{The expectation values of the thermal shift Eq. (\ref{4}), $\varepsilon_\upsilon^\beta$, in Hz at different temperatures. Levels with higher vibrational quantum number are omitted due to larger numerical error.}
\label{tab:4}
\renewcommand{\arraystretch}{1.35}
\begin{tabular}{r|c|c|c|c} 
\hline
\hline
 $\upsilon \backslash \varepsilon_\upsilon^\beta $  &  $T=70$ K & $T=200$ K & $T=300$ K & $T=400$ K \\
\hline
0 & -0.00008 & -0.00536 & -0.02725 & -0.08563 \\
1 & -0.00010 & -0.00660 & -0.03352 & -0.10691 \\
2 & -0.00012 & -0.00817 & -0.04147 & -0.13297 \\
3 & -0.00014 & -0.01018 & -0.05153 & -0.15625 \\
4 & -0.00017 & -0.01281 & -0.06450 & -0.18072 \\
5 & -0.00021 & -0.01638 & -0.08214 & -0.22493 \\
6 & -0.00029 & -0.02119 & -0.10665 & -0.31237 \\
7 & -0.00042 & -0.02774 & -0.14096 & -0.46610 \\
8 & -0.00061 & -0.03718 & -0.19000 & -0.69009 \\
9 & -0.00084 & -0.05150 & -0.26203 & -0.94992 \\
10 & -0.00108 & -0.07265 & -0.36981 & -1.19181 \\
11 & -0.00138 & -0.10644 & -0.54136 & -1.47751 \\
12 & -0.00216 & -0.16738 & -0.84901 & -2.37055 \\
13 & -0.00431 & -0.25374 & -1.47077 & -5.21058 \\
14 & -0.00844 & -0.45704 & -2.95840 & -11.17778 \\
15 & -0.01515 & -1.38407 & -6.52154 & -17.48426 \\
16 & -0.05246 & -2.82764 & -8.25021 & -11.39698 \\
\hline
\hline
\end{tabular}
\end{table}
For visual clarity, the corresponding $\varepsilon_\upsilon^\beta$ curves are displayed in Fig.~\ref{fig:10}.

The graph~\ref{fig:10} clearly shows that the thermal shift values, plotted against the vibrational quantum number $\upsilon$, change systematically with temperature. An almost exponential dependence on $\upsilon$ (linear on a logarithmic scale) is observed at all temperatures. The deviations from this law are attributed primarily to numerical errors in the highly excited vibrational levels. Although the relative inaccuracy is small, the absolute deviation is rather large, on the order of a few inverse centimeters. This deviation is reflected in the altered curve behavior for large $\upsilon$ in Fig.~\ref{fig:10}. The main finding is that the curves scale with the fourth power of temperature, as seen from their essentially parallel shifts with temperature. Therefore, no nonzero temperature (magic temperature) can make the thermal shift vanish.

\subsection{The thermal shift of the rovibrational spectrum}
\label{rVTS}

We now turn to the results for the vibrational levels, with their rotational structure included. The data for the lowest four vibrational levels for different values of the orbital angular momentum are listed in Table~\ref{tab:5}.

\begin{table}[ht!]
\caption{Rovibrational levels of the H$_2^+$ molecular ion, referenced to the $\upsilon=0, L=0$ level. For every level the lower value is calculated using the ADKB method. The upper values marked with asterisk are taken from \cite{PhysRevA.16.640}, while the other \textcolor{black}{are based on the energies (converted to cm$^{-1}$) borrowed} from \cite{Pilon_rovib}. All values are in cm$^{-1}$. Total angular momentum projection onto molecular axis was taken as $0$.}
\label{tab:5}
 \renewcommand{\arraystretch}{1.35}
\begin{tabular}{c|c|c|c|c}
\hline
\hline
$L\backslash v$                 & $0$                    & $1$             & $2$             & $3$        \\
\hline
\multirow{2}{*}{$0$}  & \multirow{2}{*}{$0.0$} & $2191.33^*$ & $4255.42^*$ & $6195.89$  \\
                      &                        & $2191.33$ & $4255.65$ & $6198.16$  \\
\hline
\multirow{2}{*}{$1$}  & $58.24^*$          & $2246.50^*$ & $4307.63^*$ & $6245.23$  \\
                      & $58.24$          & $2246.50$ & $4307.87$ & $6247.53$  \\
\hline
\multirow{2}{*}{$2$}  & $174.24^*$         & $2356.39^*$ & $4411.63^*$ & $6343.49$  \\
                      & $174.24$         & $2356.40$ & $4411.88$ & $6345.87$  \\
\hline
\multirow{2}{*}{$3$}  & $347.08$               & $2519.89$       & $4566.11$       & $6489.87$  \\
                      & $347.11$               & $2520.15$       & $4566.84$       & $6492.38$  \\
\hline
\multirow{2}{*}{$4$}  & $575.44$               & $2736.17$       & $4770.73$       & $6683.18$  \\
                      & $575.48$               & $2736.45$       & $4771.51$       & $6685.86$  \\
\hline
\multirow{2}{*}{$5$}  & $857.56$               & $3003.35$       & $5023.46$       & $6921.90$  \\
                      & $857.62$               & $3003.65$       & $5024.31$       & $6924.79$  \\
\hline
\multirow{2}{*}{$10$} & $2991.63$              & $5023.08$       & $6932.53$       & $8723.39$  \\
                      & $2991.83$              & $5023.58$       & $6934.17$       & $8728.30$  \\
\hline
\multirow{2}{*}{$15$} & $6062.80$              & $7925.48$       & $9670.94$       & $11301.64$ \\
                      & $6063.24$              & $7926.65$       & $9675.00$       & $11309.15$ \\
\hline
\multirow{2}{*}{$20$} & $9684.80$              & $11340.42$      & $12883.17$      & $14314.21$ \\
                      & $9685.74$              & $11344.01$      & $12890.90$      & $14320.62$ \\
\hline
\hline
\end{tabular}
\end{table}

In particular, Table~\ref{tab:5} presents values obtained using several different approaches~\cite{Solovyev_ADKB_2024,Pilon_rovib,PhysRevA.16.640}. Values marked with an asterisk correspond to the results of Ref.~\cite{PhysRevA.16.640}, where a numerical method for solving the Schr\"odinger equation was employed and radiative, relativistic, and nonadiabatic effects were taken into account. As a result, the accuracy of the rovibrational levels obtained in that work was established at the level of $0.002$ cm$^{-1}$. Within the ADKB approach (including the diagonal Born-Oppenheimer correction), excellent agreement with these results is found. For instance, the ground vibrational state energy given in Ref.~\cite{PhysRevA.16.640} is $-0.5971441690$ a.u., whereas the ADKB approach yields $-0.5971442779$ a.u. The relative deviation is of the order of $10^{-7}$. This may be attributed to the fact that radiative corrections, which are omitted in the ADKB method, are included in \cite{PhysRevA.16.640}. However, such a small deviation could also be due to numerical errors in our method. It should be noted that for our purposes this level of accuracy is quite sufficient. The relative deviation for the higher excited states presented in Ref.~\cite{PhysRevA.16.640} does not exceed about $10^{-6}$. \textcolor{black}{The electron-to-proton mass ratio was updated to the CODATA 2022 value (0.000 544 617 021 4889), the Ref.~\cite{codata_2022} value (0.000 544 617) yielding negligible differences for our purposes.}

Other values presented in the upper rows of Table~\ref{tab:5}, not marked with an asterisk, were taken from Ref.~\cite{Pilon_rovib}, which employed a purely nonrelativistic approach. A comparison with the results of the present work shows that the deviation, remaining below a few tenths of a percent, increases with the vibrational quantum number $\upsilon$ and reaches a few cm$^{-1}$ in absolute value at $\upsilon=3$. The energy shift for $L\neq 0$ is also in full agreement with Ref.~\cite{Ishikawa_2012}. For comparison, we also provide calculated values for larger orbital angular momenta $L=10,15,20$. The deviation from the tabulated values of Ref.~\cite{Pilon_rovib} does not exceed ten inverse centimeters, and the largest relative deviation is about $0.07\%$ for $\upsilon=3, L=15$ and $\upsilon=2, L=20$. Thus, we conservatively estimate the accuracy of the thermal shift calculation for rovibrational levels to be about $0.07\%$, which is more than sufficient for our purposes.

Having achieved such precision in the rovibrational spectrum of the hydrogen molecular ion within our fully relativistic ADKB approach, we present below the thermal shifts for the levels listed in Table~\ref{tab:5}. The shifts are evaluated by averaging the functional dependence $\Delta\varepsilon_a^\beta(R)$ over the corresponding wave functions, which are obtained simultaneously with the energy spectrum within the ADKB method. The numerical results for the diagonal matrix elements, i.e., the expectation values, are listed in Table~\ref{tab:6}. \textcolor{black}{We note that the data given in Table~\ref{tab:4} correspond to the total angular momentum with its projection onto the molecular axis taken as 1/2, while those listed in Table VI correspond to a projection of 0. This difference reflects the adiabatic separation of electronic and nuclear motion: the projection entering the centrifugal potential of the nuclear rovibrational problem is a formal parameter, which we set to an integer value when comparing with works that treat the angular momenta as integer. In a fully consistent treatment it should be half-integer, but this would require non-adiabatic corrections beyond the scope of the present paper. The two tables thus differ slightly in their values, though this does not affect the results of the present paper.}
\begin{widetext}
\begin{center}
\begin{table}[ht!]
\caption{Thermal shift, $\varepsilon^\beta_{\upsilon,L}$, for rovibrational levels of the H$_2^+$ molecular ion. The values obtained at different temperatures are given in \textcolor{black}{Hz}, the uncertainty of the presented values corresponds to a conservative estimate of $0.1\%$.}
\label{tab:6}
\renewcommand{\arraystretch}{1.35}
\resizebox{\textwidth}{!}{
\begin{tabular}{l|c|c|c|c|c|c|c|c|c}
\hline
\hline
$\upsilon$ ($70$ K) & $L=0$ & $L=1$ & $L=2$ & $L=3$ & $L=4$ & $L=5$ & $L=10$ & $L=15$ & $L=20$ \\
\hline
    0 & $-7.628\times 10^{-5}$ & $-7.652\times 10^{-5}$ & $-7.700\times 10^{-5}$ & $-7.773\times 10^{-5}$ & $-7.870\times 10^{-5}$ & $-7.994\times 10^{-5}$ & $-9.032\times 10^{-5}$ & $-1.095\times 10^{-4}$ & $-1.408\times 10^{-4}$ \\
    1 & $-9.555\times 10^{-5}$ & $-9.589\times 10^{-5}$ & $-9.658\times 10^{-5}$ & $-9.761\times 10^{-5}$ & $-9.899\times 10^{-5}$ & $-1.007\times 10^{-4}$ & $-1.150\times 10^{-4}$ & $-1.388\times 10^{-4}$ & $-1.717\times 10^{-4}$ \\
    2 & $-1.193\times 10^{-4}$ & $-1.196\times 10^{-4}$ & $-1.204\times 10^{-4}$ & $-1.216\times 10^{-4}$ & $-1.232\times 10^{-4}$ & $-1.251\times 10^{-4}$ & $-1.405\times 10^{-4}$ & $-1.656\times 10^{-4}$ & $-2.126\times 10^{-4}$ \\
    3 & $-1.425\times 10^{-4}$ & $-1.429\times 10^{-4}$ & $-1.437\times 10^{-4}$ & $-1.450\times 10^{-4}$ & $-1.466\times 10^{-4}$ & $-1.487\times 10^{-4}$ & $-1.663\times 10^{-4}$ & $-2.042\times 10^{-4}$ & $-2.974\times 10^{-4}$ \\
\hline
$\upsilon$ ($200$ K) & $L=0$ & $L=1$ & $L=2$ & $L=3$ & $L=4$ & $L=5$ & $L=10$ & $L=15$ & $L=20$ \\
\hline
    0 & $-5.357\times 10^{-3}$ & $-5.373\times 10^{-3}$ & $-5.406\times 10^{-3}$ & $-5.456\times 10^{-3}$ & $-5.522\times 10^{-3}$ & $-5.605\times 10^{-3}$ & $-6.300\times 10^{-3}$ & $-7.555\times 10^{-3}$ & $-9.618\times 10^{-3}$ \\
    1 & $-6.598\times 10^{-3}$ & $-6.620\times 10^{-3}$ & $-6.662\times 10^{-3}$ & $-6.727\times 10^{-3}$ & $-6.814\times 10^{-3}$ & $-6.924\times 10^{-3}$ & $-7.846\times 10^{-3}$ & $-9.541\times 10^{-3}$ & $-1.242\times 10^{-2}$ \\
    2 & $-8.170\times 10^{-3}$ & $-8.198\times 10^{-3}$ & $-8.254\times 10^{-3}$ & $-8.340\times 10^{-3}$ & $-8.454\times 10^{-3}$ & $-8.599\times 10^{-3}$ & $-9.828\times 10^{-3}$ & $-1.215\times 10^{-2}$ & $-1.632\times 10^{-2}$ \\
    3 & $-1.018\times 10^{-2}$ & $-1.021\times 10^{-2}$ & $-1.029\times 10^{-2}$ & $-1.040\times 10^{-2}$ & $-1.056\times 10^{-2}$ & $-1.075\times 10^{-2}$ & $-1.242\times 10^{-2}$ & $-1.572\times 10^{-2}$ & $-2.187\times 10^{-2}$ \\
\hline
$\upsilon$ ($300$ K) & $L=0$ & $L=1$ & $L=2$ & $L=3$ & $L=4$ & $L=5$ & $L=10$ & $L=15$ & $L=20$ \\
\hline
    0 & $-2.724\times 10^{-2}$ & $-2.732\times 10^{-2}$ & $-2.748\times 10^{-2}$ & $-2.773\times 10^{-2}$ & $-2.807\times 10^{-2}$ & $-2.849\times 10^{-2}$ & $-3.201\times 10^{-2}$ & $-3.835\times 10^{-2}$ & $-4.877\times 10^{-2}$ \\
    1 & $-3.351\times 10^{-2}$ & $-3.362\times 10^{-2}$ & $-3.384\times 10^{-2}$ & $-3.416\times 10^{-2}$ & $-3.460\times 10^{-2}$ & $-3.516\times 10^{-2}$ & $-3.982\times 10^{-2}$ & $-4.837\times 10^{-2}$ & $-6.275\times 10^{-2}$ \\
    2 & $-4.145\times 10^{-2}$ & $-4.159\times 10^{-2}$ & $-4.188\times 10^{-2}$ & $-4.231\times 10^{-2}$ & $-4.288\times 10^{-2}$ & $-4.361\times 10^{-2}$ & $-4.979\times 10^{-2}$ & $-6.132\times 10^{-2}$ & $-8.193\times 10^{-2}$ \\
    3 & $-5.151\times 10^{-2}$ & $-5.170\times 10^{-2}$ & $-5.207\times 10^{-2}$ & $-5.264\times 10^{-2}$ & $-5.340\times 10^{-2}$ & $-5.436\times 10^{-2}$ & $-6.264\times 10^{-2}$ & $-7.889\times 10^{-2}$ & $-1.100\times 10^{-1}$ \\
\hline
$\upsilon$ ($400$ K) & $L=0$ & $L=1$ & $L=2$ & $L=3$ & $L=4$ & $L=5$ & $L=10$ & $L=15$ & $L=20$ \\
\hline
    0 & $-8.560\times 10^{-2}$ & $-8.586\times 10^{-2}$ & $-8.637\times 10^{-2}$ & $-8.714\times 10^{-2}$ & $-8.818\times 10^{-2}$ & $-8.949\times 10^{-2}$ & $-1.006\times 10^{-1}$ & $-1.215\times 10^{-1}$ & $-1.555\times 10^{-1}$ \\
    1 & $-1.069\times 10^{-1}$ & $-1.072\times 10^{-1}$ & $-1.080\times 10^{-1}$ & $-1.091\times 10^{-1}$ & $-1.106\times 10^{-1}$ & $-1.125\times 10^{-1}$ & $-1.281\times 10^{-1}$ & $-1.533\times 10^{-1}$ & $-1.859\times 10^{-1}$ \\
    2 & $-1.329\times 10^{-1}$ & $-1.333\times 10^{-1}$ & $-1.342\times 10^{-1}$ & $-1.354\times 10^{-1}$ & $-1.370\times 10^{-1}$ & $-1.391\times 10^{-1}$ & $-1.546\times 10^{-1}$ & $-1.787\times 10^{-1}$ & $-2.251\times 10^{-1}$ \\
    3 & $-1.562\times 10^{-1}$ & $-1.566\times 10^{-1}$ & $-1.574\times 10^{-1}$ & $-1.585\times 10^{-1}$ & $-1.601\times 10^{-1}$ & $-1.620\times 10^{-1}$ & $-1.786\times 10^{-1}$ & $-2.163\times 10^{-1}$ & $-3.193\times 10^{-1}$ \\
\hline
\hline
\end{tabular}
}
\end{table}
\end{center}
\end{widetext}

The data in Table~\ref{tab:6} are well described by a cubic extrapolation in the quantum numbers $\upsilon$ and $L$: $ax^3+b x^2+c x+d$, with a fitting error of less than a few hundredths of a percent. However, the data are scalable by the factor $(k_{\rm B}T)^4$ for each $(\upsilon,L)$ only to within about $3$-$5\%$ accuracy. This behavior is attributed to the dynamical (frequency-dependent) character of the thermal shift. We therefore proceed to compare our accurately calculated thermal shifts for the rovibrational levels as obtained from the complete formula (\ref{4}) and from the approximate approach (i.e., static limit supplemented by dynamical corrections) \cite{Schiller_P}. The data are given in Table~\ref{tab:7}.






\begin{table}[ht!]
\caption{Comparison of the calculated thermal shifts for rovibrational levels of the hydrogen molecular ion at $T = 300$ K. Numerical values are taken from Table~\ref{tab:6} of the present work and Table VIII of Ref.~\cite{Schiller_P}; all results are in Hz. The last column contains the relative deviations.}
\label{tab:7}
\renewcommand{\arraystretch}{1.35}
\begin{tabular}{l|l|c|c|c}
\hline
\hline
$\upsilon$ & $L$ & $\varepsilon^\beta_{\upsilon,L}$ (this work) & $\varepsilon^\beta_{\upsilon,L}$ (see \cite{Schiller_P}) & $\delta=(\varepsilon^\beta_{\upsilon,L}-\varepsilon^{\beta \text{\cite{Schiller_P}}}_{\upsilon,L})/\varepsilon^\beta_{\upsilon,L}$ \\
\hline
0 & 0 & $-0.027236(19)$ & $-0.0273023$ & $0.00243$\\

0 & 1 & $-0.027319(19)$ & $-0.0274023$ & $0.00306$\\

0 & 3 & $-0.027734(19)$ & $-0.0278024$ & $0.00247$\\

1 & 1 & $-0.033619(24)$ & $-0.0337044$ & $0.00254$\\

1 & 3 & $-0.034164(24)$ & $-0.0342046$ & $0.00120$\\

2 & 1 & $-0.041590(29)$ & $-0.0417$ & $0.00254$\\

3 & 1 & $-0.051696(36)$ & $-0.0519156$ & $0.00423$\\
\hline
\hline
\end{tabular}
\end{table}

In Table~\ref{tab:7}, the third column contains our ADKB-based values, while the fourth column shows those from \cite{Schiller_P}. The latter results are composed of a static shift (first three significant figures) and dynamical corrections (last three figures). Despite our conservative error estimate of $0.1\%$ (e.g., $-0.02724\pm 0.00003$ for the $\upsilon=0, L=0$ state), a pronounced discrepancy persists, exceeding our numerical uncertainty by roughly an order of magnitude. The relative deviations in the last column are within $0.5\%$, whereas the relative uncertainty of our calculations is approximately $0.07\%$ (given in parentheses).

Thus, we primarily confirm the thermal shift values for the rovibrational levels of H$_2^+$ by means of a fully independent calculation. The observed discrepancies, while being largely minor, admit a straightforward interpretation. We attribute this to the static limit employed in \cite{Schiller_P}, which was supplemented by dynamical corrections to go beyond that limit. Nevertheless, an accurate integration over frequencies reveals that a full treatment of the frequency dependence in Eq.~(\ref{4}) is required; the dynamical corrections themselves do not account for the remaining difference.

\section{Conclusions}
\label{DC}

This work reports a further development of the ADKB approach for the hydrogen molecular ion, see, e.g., Ref.~\cite{Solovyev_ADKB_2024} and references therein. This approach allows us to compute the thermal shift as a function of the internuclear distance with high accuracy within a fully relativistic picture. For this purpose, we have adopted the theory of quantum electrodynamics for bound states at finite temperature, see Fig.~\ref{fig:1} and Eqs.~(\ref{4}), (\ref{5}). 

The numerical error is primarily determined by the following factors. First, by varying the size of the basis sets in the radial and azimuthal variables, we monitor the convergence of the results and their agreement with known nonrelativistic values (see the corresponding analysis in \cite{Solovyev_ADKB_2024}). Varying the box size also serves these purposes and helps to select optimal calculation parameters, including the computational time. Second, testing the results for dependence on the nuclear model used showed that this dependence is not significant for the present problem. As a next step, we monitored the convergence of the partial-wave expansion (\ref{5}): calculations were performed with $l\leqslant 3$, which far exceeds the accuracy required for the final result. In fact, the $l=2$ partial-wave contributions are already below $10^{-11}$ for all internuclear distances in our calculation (for the numerical error estimation, see Section~\ref{h2+}). The most challenging part, perhaps, is monitoring the accuracy of the frequency integration for the dynamical polarizability. To this end, we carried out calculations with different grid parameters, confirming that the corresponding numerical error is negligible. 

A major extension of the earlier ADKB version is the possibility of calculating the diagonal Born–Oppenheimer correction, which is evaluated in our calculation to an accuracy better than $10^{-3}$ of the absolute magnitude. Since this numerical uncertainty is relatively small, it has enabled us to determine the vibrational energy spectrum with an accuracy estimated at $0.04\%$, showing agreement with known results at the level of $10^{-5}$, see section~\ref{VTS} for details. Notably, the present ADKB version also includes a unified procedure for evaluating both vibrational and rovibrational levels. Consequently, our determined thermal shift values are established at the same level of accuracy.

Finally, we present calculations of the rovibrational spectrum of H$_2^+$, see section~\ref{rVTS}. Comparison with the available literature shows excellent agreement. For the full set of vibrational and rovibrational energies, we conservatively assign an uncertainty of $0.07\%$ to our thermal shift values. When compared with results from a distinctly different approach, the deviations exceed our numerical error by an order of magnitude. We believe that this discrepancy stems from the static limit and dipole approximation used by the other authors, since the dynamical corrections are much smaller than the observed deviations.

Though our main discussion of thermal shifts focuses on room temperature ($300$ K), we have also evaluated the shifts at several other temperatures for both vibrational and rovibrational levels. The calculations show that, to within about $5\%$, $T^4$ acts as a reasonable scaling factor; we attribute the $5\%$ scatter to the dynamical (frequency-dependent) nature of the thermal correction. Furthermore, for the vibrational (rovibrational) spectrum, a magic temperature (at which the thermal shift would vanish) appears unlikely to exist, see, e.g., \cite{Reiter_2026}. This circumstance underscores the importance of accurate thermal shift calculations for further improvement of precision and for the development of molecular clocks based on H$_2^+$ or its isotopologues.

Finally, the expression in Eq.~(\ref{4}) also has an imaginary part (in addition to the real part responsible for the thermal energy shift), which follows from the Sokhotski-Plemelj theorem. As shown in \cite{PRA_2015}, this imaginary part corresponds to the thermally induced level width. We have not performed such numerical evaluations here, assuming the induced widths to be negligible at room temperature. A narrow spectral line is, after all, one of the primary criteria for choosing candidate atomic or molecular species \cite{Solovyev_PP2025}.

{\it Acknowledgements.}
This work was supported by the Russian Science Foundation under grant \textnumero{25-22-00625}.
\onecolumngrid
\bibliographystyle{apsrev4-1}
\bibliography{ref} 

\end{document}